\documentclass[amsmath,amssymb,amsfonts,aps,pre,preprint,superscriptaddress,bibnotes,showpacs,showkeys,longbibliography]{revtex4-1}

\usepackage{graphicx}
\usepackage{subfigure}
\usepackage{amsmath}
\usepackage{amsfonts}
\usepackage{amssymb}
\usepackage{natbib}
\usepackage{relsize}
\usepackage{sidecap}
\usepackage{braket}
\usepackage{footmisc}
\usepackage{footnote}
\usepackage{color}
\begin{document}

\title{Capillary wave dynamics and interface structure modulation in binary Bose-Einstein condensate mixtures}
\author{Joseph O. Indekeu}
\affiliation{Institute for Theoretical Physics, KU Leuven, Celestijnenlaan 200D,  BE-3001 Leuven, Belgium}
\author{Nguyen Van Thu}
\affiliation{Department of Physics, Hanoi Pedagogical University 2, Hanoi, Vietnam}
\author{Chang-You Lin}
\affiliation{Institute for Theoretical Physics, KU Leuven, Celestijnenlaan 200D,  BE-3001 Leuven, Belgium}
\author{Tran Huu Phat}
\affiliation{Vietnam Atomic Energy Commission, 59 Ly Thuong Kiet, Hanoi, Vietnam}

\date{\today}

\begin{abstract}
The localized low-energy interfacial excitations, or Nambu-Goldstone  modes, of phase-segregated binary mixtures of Bose-Einstein condensates are investigated analytically by means of a double-parabola approximation (DPA) to the Lagrangian density in Gross-Pitaevskii theory for a system in a uniform potential. Within this model analytic expressions are obtained for the excitations underlying  capillary waves or ``ripplons" for arbitrary strength $K\,(>1)$ of the phase segregation. The dispersion relation $\omega \propto k^{3/2}$ is derived directly from the Bogoliubov-de Gennes equations in  limit that the wavelength $2\pi/k$ is much larger than the healing length $\xi $. The proportionality constant in the dispersion relation provides the static interfacial tension. A correction term in $\omega (k)$ of order $k^{5/2}$ is calculated analytically, entailing a finite-wavelength correction factor $(1+\frac{ \sqrt{K-1} \,k\xi}{4\sqrt{2}\,(\sqrt{2}+\sqrt{K-1})})$. This prediction may be tested experimentally using (quasi-)uniform optical-box traps. Explicit expressions are obtained for the structural deformation of the interface due to the passing of the capillary wave. It is found that  the amplitude of the wave is enhanced by an amount that is quadratic in the ratio of the phase velocity $\omega/k$ to the sound velocity $c$. For generic asymmetric mixtures consisting of condensates with unequal healing lengths an additional modulation is predicted of the common value of the condensate densities at the interface. 

\end{abstract}


\maketitle

\section{Introduction}

Theoretical studies of phase-segregated Bose-Einstein Condensate (BEC) mixtures and the experimental realization of such systems have opened up a new avenue for investigating many interesting properties of inhomogeneous BECs, in particular the superfluid dynamics of the interface. On the one hand, one has focused on the exploration of hydrodynamic interface instabilities, such as the Kelvin-Helmholtz instability \cite{c1}, the Rayleigh-Taylor instability \cite{c2} and the Richtmayer-Meshkov instability \cite{c3}. On the other hand, one has studied several characteristics specific to interfaces, such as the interfacial tension \cite{c4,c5} and capillary wave excitations, starting by solving approximately the Gross-Pitaevskii (GP) equations in two distinct regimes of weak and strong segregation \cite{c6,c7,c8}.

Dynamical properties of BECs are often studied numerically by solving  the time-dependent GP equations with the pseudospectral method or via diagonization of the BdG equations. Dispersion relations can to some extent be determined analytically through a variational method applied to the GP action or by solving the GP equations or BdG equations in asymptotic regimes such as the long wavelength limit. However, the analytical studies have  either been limited to special cases or have ignored the structure of the interface on length scales smaller than or equal to the healing length.

Our aim in this paper is to provide a detailed analytical solution of a model which possesses the essential physical ingredients of GP theory, but is different from it and much easier to solve. In Ref.\cite{c5} the model defined through the double-parabola approximation (DPA) was shown to be a simple but useful approach for treating {\em static} interfacial problems in phase-segregated BECs. The DPA has even played a constructive role in conjecturing new exact solutions to the GP equations \cite{c5}. The present paper proposes the {\em dynamical} counterpart of this  model. Here, the DPA in combination with the Bogoliubov-de Gennes (BdG) equations allow us to derive analytical expressions for  ripplon  excitations and  their dispersion relation. These excitations are  Nambu-Goldstone (NG) modes associated with the broken symmetries in this inhomogeneous system. Surface phonons (at an interface) can be studied in a similar fashion but are not addressed in this paper. 

A ripplon is a capillary wave traveling along the interface, causing it to undulate perpendicular to its planar projection, and is recognized as the primary lowest-energy excitation within the Bogoliubov spectrum. In a uniform external potential the dispersion relation of  ripplons, in the long-wavelength limit, $\omega^2 \propto k^3$,  where $\omega$ is the angular frequency and $k$ the wave number of the capillary wave, can be obtained by applying a linear approximation to the GP theory leading to Young-Laplace-like equations. This result coincidences with the dispersion relation of a classical fluid. The secondary low-energy excitations are phonons. While bulk phonons correspond to the propagation of the pressure waves within a condensate in bulk, localized surface phonons with dispersion  $\omega \propto k$ are predicted at the interface, propagating at a  speed lower than the speed of sound.  Drawing an analogy to mechanical waves, ripplons have an important transverse character, while surface phonons are mainly longitudinal in nature. 

Let us first recall the familiar frame-work. We assume a uniform external potential (which without loss of generality can be set to zero) and start from the GP Lagrangian density \cite{pethick2002bose,PitaString}
\begin{equation}\label{eq:lagrangiandensity}
\mathcal{L} (\Psi_1,\Psi_2 )=  \frac{i\hbar}{2}\sum_{j=1}^{2}  {\color{black} \left(\Psi_j^{\ast} \partial_t \Psi_j -\Psi_j \partial_t \Psi_j^{\ast} \right)} - \mathcal{E}(\Psi_1,\Psi_2),
\end{equation}
with Hamiltonian density
\begin{equation}\label{eq:hamiltoniandensity}
\mathcal{E} (\Psi_1,\Psi_2)=  \sum_{j=1}^{2} \left[ {\color{black} \frac{\hbar^2}{2 m_j}  \left|\nabla \Psi_j \right|^2 } + \frac{g_{jj}}{2}  \lvert\Psi_j\rvert^4  \right] + g_{12} \lvert\Psi_1\rvert^2 \lvert\Psi_2\rvert^2.
\end{equation}
For atomic species $j$, $\Psi_j=\Psi_{j}(\mathbf{r},t)$ is the wave function of the condensate or ``order parameter", $m_j$ the atomic mass, $g_{jj}=4 \pi \hbar^2 a_{jj}/m_j >0$ the repulsive intra-species interaction strength, $g_{12}=2 \pi \hbar^2 a_{12}(1/m_1 + 1/m_2)>0$ the repulsive inter-species interaction strength and $a_{jj'}$ the intra-species (for $j=j'$) or inter-species (for $j=1$ and $j'=2$) $s$-wave scattering length. 

By introducing the dimensionless quantities, $\mathbf{s}_j=\mathbf{r}/\xi_j$, with $\xi_j = \hbar/\sqrt{2m_jn_{j0}g_{jj}}$ the healing length and $n_{j0}$ the number density of condensate $j$ in bulk, $\tau_j=t/t_{j}$, $\psi_{j}=\Psi_{j}/\sqrt{n_{j0}}$, and $K=g_{12}/\sqrt{g_{11} g_{22}}$, with $t_{j}=\hbar/\mu_j$, and $\mu_j=n_{j0}g_{jj} $ the chemical potential of condensate $j$, we scale the Lagrangian density in \eqref{eq:lagrangiandensity} and Hamiltonian density in \eqref{eq:hamiltoniandensity} to
\begin{equation}\label{eq:dimensionlesslagrangiandensity}
\tilde{\mathcal{L}} (\psi_1,\psi_2) = \frac{\mathcal{L}}{2 P_0}=  \frac{i}{2} \sum_{j=1}^{2} {\color{black}\left(\psi_j^{\ast} \partial_{\tau_j} \psi_j  - \psi_j \partial_{\tau_j} \psi_j^{\ast} \right)} - \tilde{\mathcal{E}} (\psi_1,\psi_2),
\end{equation}
with
\begin{equation}
\tilde{\mathcal{E}} (\psi_1,\psi_2)= \frac{\mathcal{E}}{2 P_0} = \sum_{j=1}^{2} \left[ {\color{black} \left| \nabla_{\mathbf{s}_j} \psi_j \right|^2 }+ \frac{\lvert\psi_j\rvert^4}{2} \right] + K \lvert\psi_1\rvert^2 \lvert\psi_2\rvert^2,
\end{equation}
where the bulk pressure $P_0$ is given by $\mu_j^2/2 g_{jj}$, which is independent of the label $j$ provided the mixture is at bulk two-phase coexistence, which indeed we must assume in order to ensure the stability of an interface in bulk. 
Next we make a transformation of the dimensionless Lagrangian density by writing 
\begin{equation}
\psi_j (\mathbf{s}_j,\tau_j) \equiv \phi_j (\mathbf{s}_j,\tau_j)e^{- i\tau_j }.
\end{equation}
We then have a Lagrangian density in terms of the new order parameters $\phi_j$, 
\begin{equation}\label{eq:dimensionlesslagrangiandensityinphi}
\hat{\mathcal{L}} \left(\phi_1,\phi_2 \right) \equiv \tilde{\mathcal{L}} \left(\phi_1 e^{- i\tau_1}, \phi_2 e^{- i\tau_2} \right) = \sum_{j=1}^{2}  \left[  {\color{black} \frac{i}{2} \left( \phi_j^{\ast}  \partial_{\tau_j} \phi_j - \phi_j \partial_{\tau_j} \phi_j^{\ast} \right) - \left|\nabla_{\mathbf{s}_j} \phi_j \right|^2 } \right] - \hat{\mathcal{V}} (\phi_1,\phi_2),
\end{equation}
in which the potential $\tilde{\mathcal{V}}$ takes the form
\begin{equation}\label{eq:potential}
\tilde{\mathcal{V}}(\phi_1,\phi_2)= \sum_{j=1}^{2}  \left[-\lvert\phi_j\rvert^2 + \frac{\lvert\phi_j\rvert^4}{2} \right] + K \lvert\phi_1\rvert^2 \lvert\phi_2\rvert^2.
\end{equation}
Recall that for $K>1$ the two components are immiscible and a phase-segregated BEC forms \cite{ao}. We assume that the planar interface (without excitation) is located at $z = 0$ and the condensates 1 and 2 reside in half-spaces $z\geq0$ and $z\leq0$, respectively.

We call the BEC mixture {\em symmetric} when the atomic masses are equal, $m_1=m_2$ (a frequently fulfilled condition), {\em and} the intra-species interaction strengths are equal $g_{11}=g_{22}$ (a special requirement). Furthermore, for a symmetric mixture at bulk two-phase coexistence also the bulk densities are  equal, $n_{10}=n_{20}$, and so are the chemical potentials. Consequently, $\xi_1= \xi_2$ for a symmetric mixture.  

\section{DPA applied within  GP theory}

\subsection{DPA for the GP equations}

We recall the dimensionless mean-field DPA GP Lagrangian density \cite{c5}
\begin{equation}\label{eq:DPALagrangiandensity}
\hat{\mathcal{L}} \left(\phi_1,\phi_2 \right) = \sum_{j=1}^{2}  \left[  \frac{i}{2} \left( \phi_j^{\ast}  \partial_{\tau_j} \phi_j - \phi_j \partial_{\tau_j} \phi_j^{\ast} \right) - \left|\nabla_{\mathbf{s}_j} \phi_j \right|^2  \right] - \hat{\mathcal{V}}_{\mathrm{DPA}}(\phi_1,\phi_2),
\end{equation}
where $\hat{\mathcal{V}}_{\mathrm{DPA}}$ takes the form
\begin{equation}\label{eq:DPApotential}
\hat{\mathcal{V}}_{\mathrm{DPA}}(\phi_1,\phi_2) = 2 \left( \left|\phi_j \right| -1\right)^2  + \beta^2 \lvert \phi_{j'} \rvert^2 - \frac{1}{2},
\end{equation}
where $\beta=\sqrt{K-1}$. In \eqref{eq:DPApotential} the labels $j$ and $j'$ comply with the following important convention, which we will henceforth maintain {\em throughout this paper} :
\begin{equation}\label{cases}
 (j,j')=
\begin{cases}
(1,2) & \text{for} \; z \ge 0 \\
(2,1) & \text{for} \; z \le 0,
\end{cases}
\end{equation}
This potential is the combination of two paraboloids obtained by the expansions of the potential in GP theory about bulk condensate 1, $(\left|\phi_1\right|,\left|\phi_2\right|) =(1,0)$, for the half-space $z \ge 0$, and bulk condensate 2, $(\left|\phi_1\right|,\left|\phi_2\right|) = (0,1)$, for the half-space $z \le 0$. In bulk, time derivatives and spatial gradients in the Lagrangian density \eqref{eq:DPALagrangiandensity} vanish and the minimum of the potential, $-1/2$, is the same as in the GP theory, see \cite{c5}.

By extremizing the action with the Lagrangian density \eqref{eq:DPALagrangiandensity} with respect to a variation  $\delta\phi_j^{\ast}$, we obtain the  DPA to the time-dependent GP equations, 
\begin{eqnarray}\label{eq:DPAGPE}
\begin{split}
i \partial_{\tau_j} \phi_j & = - \nabla_{\mathbf{s}_j}^2 \phi_j + 2 \left( \phi_j -e^{i \theta_j}\right) \\&= - \nabla_{\mathbf{s}_j}^2 \phi_j + 2 \left( |\phi_j| -1 \right) e^{i \theta_j}\\
i \partial_{\tau_{j'}} \phi_{j'} & = - \nabla_{\mathbf{s}_{j'}}^2 \phi_{j'} + \beta^2 \phi_{j'},
\end{split}
\end{eqnarray}
where $\phi_j = \left| \phi_j\right| \exp(i\theta_j)$. These equations are decoupled, which greatly simplifies the problem. Although most terms are linear in the order parameters, there is an important nonlinearity through the nontrivial dependence on the phase of the order parameter. The equations can also be obtained directly from the (exact) GP equations by expanding the amplitudes to first order in the perturbations about their bulk values. To see this, we make use of the original expansions that define the DPA \cite{c5} 
\begin{equation}\label{eq:DPAphij}
|\phi_{j} | =1+\epsilon_j; \; |\phi_{j'}| = \delta_{j'},
\end{equation}
and rewrite the (exact) GP equations, given in \eqref{eq:GPE} in Appendix A, in the form,
\begin{equation}\label{eq:DPAonGPE}
\begin{split}
i \partial_{\tau_j} \phi_j & = \left (- \nabla_{\mathbf{s}_j}^2 -1  + (1+\epsilon_j)^2 + K \delta_{j'} ^2 \right ) (1+\epsilon_j) e^{i\theta_j} \\&\approx - \nabla_{\mathbf{s}_j}^2 (1+\epsilon_j) e^{i\theta_j}+ 2\epsilon_j e^{i\theta_j}\\ 
i \partial_{\tau_{j'}} \phi_{j'} & = \left (- \nabla_{\mathbf{s}_{j'}}^2 -1 + \delta_{j'}^2 + K(1+\epsilon_j)^2\right )\delta_{j'}e^{i\theta_{j'}}\\ & \approx (- \nabla_{\mathbf{s}_{j'}}^2 +K -1 ) \delta_{j'}e^{i\theta_{j'}}
\end{split} 
\end{equation}
We observe that keeping contributions of zeroth and first order in the $\epsilon$'s and $\delta$'s reproduces \eqref{eq:DPAGPE} identically.
Alternatively, one can derive these equations ``hydrodynamically" by combining a Bernoulli-type of equation and the continuity equation, obtained by taking the variation of the amplitudes and the phases of the order parameters, respectively \cite{Stringari}. 

For a planar interface at $z=0$, the static interface profiles are real solutions of the form of time-independent stationary states $\phi_j (\rho_j,\tau_j) = \phi_{j0} (\rho_j)$, with $\rho_j \equiv z/\xi_j$, while the (unperturbed) system is translationally invariant along $x$ and $y$ directions. By matching the functions and their first derivatives at $z=0$ and  using the boundary conditions $\phi_{10}(\infty) =\phi_{20}(-\infty)=1$ and $\phi_{20}(\infty) =\phi_{10}(-\infty) =0$, we obtain
\begin{equation}\label{eq:stationarysolutionsDPAGPE}
\begin{split}
\phi_{j0} (\rho_j) & = 1 - \frac{\beta}{\sqrt{2}\, + \beta} e^{- \sqrt{2}\, \left| \rho_j \right| } \\
\phi_{j'0} (\rho_{j'}) & = \frac{\sqrt{2}\,}{\sqrt{2}\, + \beta}e^{- \beta \left| \rho_{j'} \right| } 
\end{split}
\end{equation}
Note that the second derivative of these profiles is discontinuous at the interface.

After inserting the real and time-independent solutions \eqref{eq:stationarysolutionsDPAGPE} in  \eqref{eq:DPAGPE}, we multiply the equation for condensate $j$ by $\partial_z \phi_{j0}$  and the equation  for condensate $j'$ by $\partial_z \phi_{j'0}$, then add them up and integrate the expression over $z$. The result is a constant and represents the constant of motion in DPA, i.e.,
\begin{equation}
\sum_{j=1}^{2} \left(\partial_{\rho_j} \phi_{j0} \right)^2 +\hat{\mathcal{V}}_{\mathrm{DPA}}(\phi_{10},\phi_{20}) = -\frac{1}{2}.
\end{equation}
The value of the constant must be fixed so that the minimum of  $\hat{\mathcal{V}}_{\mathrm{DPA}}$, given in \eqref{eq:DPApotential}, is respected in each of the (homogeneous) bulk phases. Note that the first term in the LHS of this equation is the sum of gradients squared of the two stationary solutions and is usually called the gradient energy.

\subsection{DPA  for the BdG equations}

Considering low-energy interfacial excitations such as ripplons and surface phonons, or even more general excitations, we perturb the complex order parameters about their (real) stationary state forms
\begin{equation}\label{eq:Bogoliubovdeltaphi}
\phi_j (\mathbf{s}_j,\tau_j) = \phi_{j0} (\mathbf{s}_j) + \delta \phi_j (\mathbf{s}_j,\tau_j),
\end{equation}
in which $|\delta \phi_j|$ is assumed to be a small perturbation of the value of $\phi_{j0}$. This implies that the phase angle $\theta_j$ of $\phi_j$ is small (i.e., of the order of Im$(\delta \phi_j)$) in regions where $\phi_{j0}$ is close to its bulk value (which is unity). In GP theory, by taking this perturbation into the GP equations, we can derive the BdG equations to first order in $\delta \phi_j$. Alternatively, the BdG equations can be derived by including the variation $\delta \phi_j^{\ast}$ directly in the GP action, using \eqref{eq:Bogoliubovdeltaphi}. For a brief recapitulation of these fundamental equations, see Appendix~\ref{sec:GPtheory}.

Within the DPA, however, a considerable simplification of these BdG equations can be obtained, while keeping -- as we shall show -- the essential physics of our problem intact. We derive the DPA to the BdG equations directly as follows. Firstly, we insert the perturbed order parameter \eqref{eq:Bogoliubovdeltaphi} into the time-dependent DPA to the GP equations \eqref{eq:DPAGPE}. This leads to 
\begin{equation}\label{eq:DPABdGEdeltaphi1}
\begin{split}
i \partial_{\tau_j} \delta\phi_j & = -\nabla_{\mathbf{s}_j}^2 \delta\phi_j + 2\left(\delta\phi_j + 1-e^{i \theta_j} \right)\\
i \partial_{\tau_j'} \delta\phi_{j'} & = - \nabla_{\mathbf{s}_{j'}}^2 \delta\phi_{j'} + \beta^2\delta\phi_{j'}
\end{split}
\end{equation}
Next, this can be rewritten in terms of the perturbations alone if we make the additional assumption that the phase angle $\theta_j$ is small. This is valid provided $|\delta \phi_j| \ll \phi_{j0}$. Since condensate $j$ is close to its bulk density in the half-space concerned, this assumption is appropriate within the DPA strategy.  Then, to first order in the $\delta \phi_j$,
\begin{equation}
e^{i \theta_j} -1 \approx i \sin \theta_j\approx \frac{\phi_j - \phi_j^{\ast}}{2|\phi_j|} \approx \frac{\delta\phi_j - \delta\phi_j^{\ast}}{2\phi_{j0}}\approx (1-\epsilon_{j0})\frac{\delta\phi_j - \delta\phi_j^{\ast}}{2},
\end{equation}
with the definition
\begin{equation}\label{eq:DPAphij0}
\phi_{j0} = 1 + \epsilon_{j0}
\end{equation}
and $\epsilon_{j0}$ is considered small (compared to 1) in the usual DPA spirit. This is a good approximation away from the interface, whereas close to the interface the upper bound is $|\epsilon_{j0}| \leq \sqrt{K-1}/(\sqrt{2}+ \sqrt{K-1}) < 1$, as can be seen from \eqref{eq:stationarysolutionsDPAGPE}. 

Consequently, \eqref{eq:DPABdGEdeltaphi1} can be written in the more canonical form, 
\begin{equation}\label{eq:DPABdGEdeltaphi}
\begin{split}
i \partial_{\tau_j} \delta\phi_j & = -\nabla_{\mathbf{s}_j}^2 \delta\phi_j + \delta\phi_j + \delta\phi_j^{\ast} + \epsilon_{j0}(\delta\phi_j - \delta\phi_j^{\ast})\\
i \partial_{\tau_j'} \delta\phi_{j'} & = - \nabla_{\mathbf{s}_{j'}}^2 \delta\phi_{j'} + \beta^2\delta\phi_{j'}
\end{split}
\end{equation}
which {\em defines} our DPA model at the level of the BdG equations. We will see shortly that the correction term $\epsilon_{j0}(\delta\phi_j - \delta\phi_j^{\ast})$ in these equations is crucially important to obtain the correct physics (including the correct ``zero modes") within the DPA. 

For completeness we also give the explicit expressions for the modulus and the phase angle of the perturbed wave function to the lowest relevant orders in the perturbation and in the deviations from bulk values,
\begin{eqnarray}
\epsilon_j &= &\epsilon_{j0} + {\rm Re}(\delta \phi_j)\\
\theta_j &=& (1-\epsilon_{j0}) {\rm Im}(\delta \phi_j).
\end{eqnarray}
This shows that ${\rm Im}(\delta \phi_j)$ is of first order in $\theta_j$, but of zeroth order in the $\epsilon$, whereas ${\rm Re}(\delta \phi_j)$ is of first order in the $\epsilon$ (or smaller). 
We conclude that  terms of order $\epsilon_{j0}{\rm Re}(\delta \phi_j)$ can be neglected, whereas terms of order $\epsilon_{j0} {\rm Im}(\delta \phi_j)$ ought to be kept.

Based on these insights we make the observation that the DPA procedure commutes with the derivation of the BdG equations. Indeed, starting from the exact BdG equations (see Appendix~\ref{sec:GPtheory}) and applying  \eqref{eq:DPAphij} and \eqref{eq:DPAphij0} also leads to  \eqref{eq:DPABdGEdeltaphi} as we will now show. The exact BdG equations, derived from GP theory, read
\begin{equation}\label{eq:BdGEdeltaphiMain}
i \partial_{\tau_j} \delta \phi_j  = \left[- \nabla_{\mathbf{s}_j}^2 -1 + 2 \phi_{j0}^2 + K \phi_{j'0}^2 \right] \delta\phi_j + \phi_{j0}^2 \delta\phi_j^{\ast} + K \phi_{j0}\phi_{j'0} (\delta\phi_{j'} +\delta\phi_{j'}^{\ast}), \; j=1,2 \; (j \ne j')
\end{equation}
If we consider the half-space in which condensate $j$ is close to its bulk state, we can apply \eqref{eq:DPAphij} and \eqref{eq:DPAphij0}. Applying the DPA we can then safely neglect the second-order quantities $\phi_{j'0}^2$ and $\phi_{j'0} (\delta\phi_{j'} +\delta\phi_{j'}^{\ast})$. This already reduces the equation to 
\begin{equation}\label{eq:BdGEdeltaphiMainRedu}
i \partial_{\tau_j} \delta \phi_j  = \left[- \nabla_{\mathbf{s}_j}^2 -1 + 2 \phi_{j0}^2 \right] \delta\phi_j + \phi_{j0}^2 \delta\phi_j^{\ast}, \; \mbox{with} \; \phi_{j0} = 1 +\epsilon_{j0}
\end{equation}
Next we apply \eqref{eq:DPAphij0} and expand, to obtain
\begin{equation}\label{eq:BdGEdeltaphiMainReduNext}
i \partial_{\tau_j} \delta \phi_j  = \left[- \nabla_{\mathbf{s}_j}^2 -1 + 2 (1+2\epsilon_{j0}) \right] \delta\phi_j + (1+2\epsilon_{j0}) \delta\phi_j^{\ast}
\end{equation}
Subsequently we regroup terms and neglect $\epsilon_{j0}{\rm Re}(\delta \phi_j)$ but keep $\epsilon_{j0} {\rm Im}(\delta \phi_j)$ as we have explained. This leads to 
\begin{equation}\label{eq:BdGEdeltaphiMainReduNext}
i \partial_{\tau_j} \delta \phi_j  = \left[- \nabla_{\mathbf{s}_j}^2 +1 \right] \delta\phi_j + \delta\phi_j^{\ast}+\epsilon_{j0} ( \delta\phi_j - \delta\phi_j^{\ast}), 
\end{equation}
which is identical to the first equation in \eqref{eq:DPABdGEdeltaphi}. Conversely, considering the half space in which condensate $j'$ is close to its bulk state, prompts us to neglect the second-order quantities $\phi_{j0}^2$ and $\phi_{j0}(\delta\phi_{j'} +\delta\phi_{j'}^{\ast})$, so that only the following terms remain,
\begin{equation}\label{eq:BdGEdeltaphiMainConv}
i \partial_{\tau_j} \delta \phi_j  = \left[- \nabla_{\mathbf{s}_j}^2 -1 +  K \phi_{j'0}^2 \right] \delta\phi_j
\end{equation}
After writing $\phi_{j'0} = 1 +\epsilon_{j'0}$ and keeping only the leading order, since $\delta \phi_j$ is already of first order in the DPA, this equation reproduces the second equation in \eqref{eq:DPABdGEdeltaphi}, if we take into account the labeling 
convention \eqref{cases}. This completes the alternative derivation of \eqref{eq:DPABdGEdeltaphi}.

Following the Bogoliubov analysis, we now write a general $\delta \phi_j$ as a linear superposition of different Fourier-like modes
\begin{equation}\label{eq:deltaphi}
\delta \phi_j (\mathbf{s}_j,\tau_j) = \sum_\mathbf{k} \left[ u_{j\mathbf{k}}  (\mathbf{s}_j) e^{i (\hat{\mathbf{k}}_j \cdot \mathbf{s}_j - \hat{\omega}_j \tau_j)} + v_{j\mathbf{k}}^{\ast} (\mathbf{s}_j) e^{-i (\hat{\mathbf{k}}_j \cdot \mathbf{s}_j - \hat{\omega}_j \tau_j)} \right],
\end{equation}
where the reduced wave vector $\hat{\mathbf{k}}_j$ and reduced frequency $\hat\omega_j$ for each mode of the surface wave are defined by $\hat{\mathbf{k}}_j=\mathbf{k}\,\xi_j$ and $\hat{\omega}_j=\hbar \omega/\mu_j=\omega t_{j}$. Note that $\hat{\mathbf{k}}_j \cdot \mathbf{s}_j - \hat{\omega}_j \tau_j = \mathbf{k}\cdot \mathbf{r} - \omega t$. Note also that for a symmetric BEC mixture at two-phase coexistence the reduced frequencies $\hat{\omega}_j$ are the same for both condensates.

\section{Ripplon-like excitations}
\subsection{Ripplon representation in GP theory}

For an interfacial excitation (a ripplon or, also, a localized phonon) propagating along a direction, say $x$, parallel to the topological defect (i.e., the interface midplane), the problem simplifies to a two-dimensional one, with $\chi_j \equiv x/\xi_j$,
\begin{equation}\label{eq:perturbedphi1d}
\phi_j (\rho_j, \chi_j, \tau_j) = \phi_{j0} (\rho_j) + \delta \phi_j (\rho_j, \chi_j, \tau_j),
\end{equation}
and
\begin{equation}\label{eq:deltaphiuv}
\delta \phi_j (\rho_j, \chi_j, \tau_j) = \sum_k \left[ u_{jk}  (\rho_j) e^{i (\hat{k}_j \chi_j - \hat{\omega}_j \tau_j)} + v_{jk}^{\ast} (\rho_j) e^{-i (\hat{k}_j \chi_j- \hat{\omega}_j \tau_j)} \right],
\end{equation}

We define the interface displacement $\zeta(x,t)$ and its dimensionless counterparts $\varrho_j (\chi_j,\tau_j) \equiv \zeta(x,t)/\xi_j$ through the position along $z$ where the order parameters ``intersect",
\begin{equation}\label{eq:intersection}
\left| \phi_1(\rho_1 = \varrho_1,\chi_1, \tau_1)\right|= \left|\phi_2 (\rho_2=\varrho_2,\chi_2, \tau_2) \right|
\end{equation}
Note that $\zeta = \varrho_j =0$ in the absence of an excitation. The converse need not be true. For example, interface breather modes are excitations for which we can fix $\zeta = \varrho_j =0$ for all times. For a ripplon mode,
the interface displacement is a transverse wave traveling along $x$.

We consider a ripplon mode of fixed wavelength and study its properties. For concreteness, let
\begin{equation}\label{eq:interfacefluctuation}
\varrho_j(\chi_j,\tau_j)  = \varrho_{j0} \sin(\hat{k}_j \chi_j - \hat{\omega}_j \tau_j) = (\zeta_0/\xi_j) \sin(kx-\omega t),
\end{equation}
where $\zeta_0$ is the amplitude of the interface displacement $\zeta$. Importantly, we assume the excitation is small in the sense that $\varrho_{j0} \ll 1$. This means that the healing lengths are large compared to the interface displacement and we anticipate that interface structure plays an important physical role in this regime.

For the order parameters associated with this ripplon the following form is proposed, 
\begin{equation}\label{eq:ansatz}
\phi_j(\rho_j,\chi_j,\tau_j)=\phi_{j0} \left (\rho_j-\varrho_{j}^{(R)}(\chi_j,\tau_j) \right)e^{-\eta_j(\rho_j,\chi_j,\tau_j)}e^{i\theta_j(\rho_j,\chi_j,\tau_j)},
\end{equation}
featuring the rigid-shift displacement $\zeta^{(R)}(x,t)$ and its dimensionless counterparts $\varrho_{j}^{(R)} (\chi_j,\tau_j) \equiv \zeta^{(R)}(x,t)/\xi_j$. These are of the same form as  \eqref{eq:interfacefluctuation}, but the amplitudes may differ, i.e.,
\begin{equation}\label{eq:interfacefluctuationRIGID}
\varrho_{j}^{(R)}(\chi_j,\tau_j)  = \varrho_{j0}^{(R)} \sin(\hat{k}_j \chi_j - \hat{\omega}_j \tau_j) = (\zeta_{0}^{(R)}/\xi_j) \sin(kx-\omega t)
\end{equation}
 Further, 
the real function
 $\eta_j$ incorporates amplitude fluctuations beyond the bare rigid shift of the static profile already incorporated in the argument of $\phi_{j0} $. The presence of  $\eta_j$ can cause a (condensate-dependent) additional shift and also a deformation of the interface and this is the physical reason for a possible difference between the functions $\zeta^{(R)}$ and $\zeta$. Finally, the real function $\theta_j$ contains the phase fluctuations (relative to $\theta =0$). For the specific interface displacement \eqref{eq:interfacefluctuation} consistency of the transverse interface displacement velocity with the phase fluctuation of each component,
\begin{equation}\label{vconsistent}
v_{jz} \equiv \frac{\hbar}{m_j} (\nabla \theta_j)_z =\partial _t \zeta,
\end{equation}
and satisfaction of the hydrodynamic equation of continuity, for each component,
\begin{equation}\label{conteq}
\partial_t |\psi_ j | ^2 + {\bf \nabla} . (|\psi_ j | ^2 {\bf v}_j) =0
\end{equation}
 imply the following forms to first order in the rigid-shift displacement amplitude $\zeta_0^{(R)}$,
\begin{equation} \label{etaandtheta}
\begin{split}
\eta_j &=  \varrho_{j0}^{(R)}  F_j (\rho_j) \sin(\hat{k}_j \chi_j -  \hat{\omega}_j \tau_j) \\
\theta_j &=  \varrho_{j0}^{(R)}   f_j (\rho_j) \cos(\hat{k}_j \chi_j -  \hat{\omega}_j \tau_j),
\end{split}
\end{equation}
where $F_j$ and $f_j$ are (real) functions that are localized about the interface, since a ripplon is a localised NG mode, whose effect decays when we move away from the interface into the bulk. These functions are related to one another through the continuity equation \eqref{conteq}, which, after scaling the variables, takes the form
\begin{equation}\label{continuity}
\hat \omega_j (\partial_{\rho_j} \phi_{j0} +\phi_{j0}F_j) = \hat k_j^2\phi_{j0} f_j   -2 (\partial_{\rho_j} \phi_{j0}) \partial_{\rho_j} f_j - \phi_{j0} \partial_{\rho_j}^2 f_j = \hat k_j^2\phi_{j0} f_j  - \partial_{\rho_j}^2 (\phi_{j0} f_j) + f_j\partial_{\rho_j}^2 \phi_{j0},
\end{equation}
to first order in the amplitude $\zeta_0^{(R)}$.
These yet unknown functions $F_j$ and $f_j$, which in general also depend on $\hat{k}_j$ and $\hat{\omega}_j$, contain the interesting physics of the excitation. For $F_j =0$ the interface is merely rigidly shifted with respect to its unperturbed state. For $F_j \neq 0$ the interface structure may undergo a change.  Further, the properties and form of $f_j $ are of course crucial  for the existence of a superfluid velocity profile. Note that far away from the interface (in the ``far field"), where condensate $j$ is present in bulk, we can approximate $\phi_{j0} \approx 1$ and the equation of continuity simplifies to
\begin{equation}\label{conteqfar}
\hat \omega_j F_j -   \hat k_j^2 f_j + \partial_{\rho_j}^2 f_j \approx 0,\; \mbox{in the far field}
\end{equation}

We now expand $\phi_j$ in \eqref{eq:ansatz}  to first order in $\varrho_{j0}^{(R)}$ and match it with $\phi_j$ in \eqref{eq:perturbedphi1d} using the term corresponding to wave number $k$ in $\delta \phi_j$ given in \eqref{eq:deltaphiuv}. This leads to the following results 
\begin{equation}\label{generalformDS}
\begin{split}
\Sigma_{jk} & \equiv u_{jk} + v_{jk} = i \varrho_{j0}^{(R)}\left ( \partial_{\rho_j} \phi_{j0} + \phi_{j0}   F_{jk}\right) \\
\Delta_{jk} & \equiv u_{jk} - v_{jk} =  i \varrho_{j0}^{(R)} \,\phi_{j0} \,f_{jk},
\end{split}
\end{equation}
where we have included the subscript $k$ in the functions $F$ and $f$ to anticipate their explicit dependence on the wave number. 
Note that the separation of the derivative part and the part containing $F_{jk}$ in the equation for $\Sigma_{jk} $ is not unique and depends on the precise form of the proposed order parameter function \eqref{eq:ansatz}. In contrast, the expression for $\Delta_{jk} $ is generally valid, since it is merely a tautology featuring the yet unknown function $f_{jk}$.

The continuity conditions, at the interface ($z=0$), for these functions and their first derivatives are a nontrivial problem in general. We will see in the next subsection how these conditions naturally manifest themselves in the DPA model. We can already anticipate that a discontinuous second derivative in $\phi_{j0}$ at the interface, typical for DPA profiles, generates a discontinuous first derivative in $\Sigma_{jk}$. In fact, even a discontinuity in $\Sigma_{jk}$ at the interface is possible, and this seems to arise naturally in the strong segregation limit, in which the exact GP interface profile (as well as its DPA counterpart) displays a discontinuous first derivative. We conclude that continuity conditions on $\Sigma_{jk}$ have to be examined carefully case by case. Since the structure of the interface is by and large determined by the properties of $\phi_{j0}$ and its first derivative, it is apt to denote the boundary conditions on $\Sigma_{jk}$ by ``structural" boundary conditions.

A different situation arises when it comes to $\Delta_{jk}$. In view of \eqref{etaandtheta}, the function $f_{jk}$ is  entangled with the phase that defines the velocity potential of the superfluid. The continuity, at the interface, of the first derivative of $f_{jk}$, is a prerequisite for the fluid mechanical continuity of the transverse velocity of the interface displacement. Therefore, we may denote the requirement of continuity of the first derivative of $\Delta_{jk}$ at the interface by ``hydrodynamical" boundary condition. Naturally, we require $\Delta_{jk}$ itself to be continuous also (since a  delta-function-like singularity in the transverse velocity would be unacceptable).

\subsection{Mathematical DPA solutions for ripplons}

By substituting $\delta \phi_j $ with the Bogoliubov form corresponding to fixed wave number $k$ in \eqref{eq:deltaphiuv} into the DPA for the BdG equations \eqref{eq:DPABdGEdeltaphi}, we obtain the following DPA for the BdG equations in terms of  $\Sigma_{jk}$ and $\Delta_{jk}$, 
\begin{equation}\label{eq:DPABdGESigmaDelta}
\begin{split}
\left( \mathcal{A}_{jk} + 2\right) \Sigma_{jk} &= \hat{\omega}_j \Delta_{jk} \\
 \mathcal{A}_{jk}  \Delta_{jk} + 2(\phi_{j0} -1)\Delta_{jk}&= \hat{\omega}_j \Sigma_{jk} \\
\left( \mathcal{A}_{j'k} + \beta^2 \right)  \Sigma_{j'k} &= \hat{\omega}_{j'} \Delta_{j'k} \\
\left( \mathcal{A}_{j'k} + \beta^2 \right) \Delta_{j'k} &= \hat{\omega}_{j'} \Sigma_{j'k}
\end{split}
\end{equation}
where $\mathcal{A}_{jk} = \hat{k}_j^2- \partial^2_{\rho_j}$. 

To test the usefulness of these equations, we first apply them to obtain the zero modes within DPA. In the limit $\hat k \rightarrow 0$ and $\hat \omega \rightarrow 0$ the DPA BdG equations \eqref{eq:DPABdGESigmaDelta} reduce to
\begin{equation}\label{eq:DPABdGESigmaDelta0}
\begin{split}
\left( - \partial^2_{\rho_j} + 2\right) \Sigma_{j0} &= 0\\
\left( - \partial^2_{\rho_j} +2(\phi_{j0} -1)\right)\Delta_{j0}&= 0 \\
\left( - \partial^2_{\rho_j} + \beta^2 \right)  \Sigma_{j'0} &= 0 \\
\left( - \partial^2_{\rho_j}+ \beta^2 \right) \Delta_{j'0} &= 0
\end{split}
\end{equation}

We now show that these equations are solved by the functions that are precisely the DPA applied to the exact zero modes of the BdG equations. To see this, consider the exact zero modes given in \eqref{exactzeroD} and \eqref{exactzeroS} of Appendix A. Then write, as usual in the DPA formalism,
\begin{eqnarray}\label{expand}
\begin{split}
\phi_{j0} &= 1+\epsilon_{j0}\\
\phi_{j'0} &= \delta_{j'0},
\end{split}
\end{eqnarray}
where the expressions for the supposedly small (compared to unity) quantities $\epsilon_{j0}$ and $\delta_{j'0}$ can be read off from \eqref{eq:stationarysolutionsDPAGPE}. At the interface $\epsilon_{j0}$ and $\delta_{j'0}$ cannot both be small, but fortunately they are well bounded in absolute value by $\beta/(\sqrt{2}+\beta)$ and $\sqrt{2}/(\sqrt{2}+\beta)$, respectively. Now recall that in DPA calculations we neglect contributions of second and higher order in the $\epsilon$'s and $\delta$'s. We concentrate first on the nontrivial second equation of \eqref{eq:DPABdGESigmaDelta0}, and test a solution of the form $\phi_{j0}$. We get
\begin{equation}\label{testsolve}
\left( - \partial^2_{\rho_j} +2\epsilon_{j0} \right)(1+\epsilon_{j0})\approx  - \partial^2_{\rho_j} \epsilon_{j0} +2\epsilon_{j0} = 0,
\end{equation}
where we have neglected the term of second order in $\epsilon_{j0}$.
The remaining three equations are trivial to test and the solutions are easily identified. Altogether, we find that the DPA BdG equations are solved by zero modes of the form 
\begin{equation}\label{DPAzeroS}
\Sigma_{j0} \propto \partial_{\rho_j} \phi_{j0}, \;\; j=1,2,
\end{equation}
and
\begin{equation}\label{DPAzeroD}
\Delta_{j0} \propto \phi_{j0}, \;\; j=1,2,
\end{equation}
provided the DPA solutions \eqref{expand} of the static order parameters are used and calculations are carried out to, and including, first-order terms in the deviations $\epsilon$ and $\delta$ from bulk values.

Returning now to the non-zero modes, we emphasize that the nontrivial second differential equation of \eqref{eq:DPABdGESigmaDelta} embodies the equation of continuity \eqref{continuity}, to first order in $\epsilon_{j0}$. This can be seen by using the forms \eqref{generalformDS} and by using also $\partial_{\rho_j}^2 \phi_{j0}  = 2 (\phi_{j0}-1)$, which is valid to first order in $\epsilon_{j0}$, as follows from \eqref{eq:DPAGPE} for a static solution. The second equation of \eqref{eq:DPABdGESigmaDelta} therefore represents {\em the DPA to the continuity equation} and is valid for all distances, at and away from the interface.

To make further progress towards the general solutions of the DPA BdG equations, we recall that the correction term in the second equation of \eqref{eq:DPABdGESigmaDelta} contains the factor $\phi_{j0}-1$, which is itself already a first-order correction in the DPA strategy. Consequently, here it is sufficient to take into account only the contribution to $\Delta_{jk}$ calculated in the absence of the correction term. We will deal with this systematically in the frame-work of perturbation theory.

Firstly, let us  denote by $\Delta_{jk}^{(0)}$ and $\Sigma_{jk}^{(0)}$ the ``zeroth-order" solutions of the  DPA BdG equations in the absence of the term $2(\phi_{j0} -1)\Delta_{jk}$ in \eqref{eq:DPABdGESigmaDelta}. These equations are solved by the exponentially localized functions
\begin{equation}\label{eq:DPABdGESigmaDeltasolution}
\begin{split}
\Sigma^{(0)}_{jk} &= A_{jk+} e^{-  \alpha_{j+} \left| \rho_j \right|} + A_{jk-} e^{- \alpha_{j-} \left| \rho_j \right|}  \\
\Delta^{(0)}_{jk} &= B_{jk+} e^{-   \alpha_{j+} \left| \rho_j \right|} + B_{jk-} e^{-  \alpha_{j-} \left| \rho_j \right|}  \\
\Sigma_{j'k} &= C_{j'k+} e^{-  \beta_{j'+}  \left|\rho_{j'} \right|} + C_{j'k-} e^{-  \beta_{j'-} \left| \rho_{j'} \right|}\\
\Delta_{j'k} &= D_{j'k+} e^{-  \beta_{j'+} \left| \rho_{j'} \right|} + D_{j'k-} e^{-  \beta_{j'-} \left| \rho_{j'} \right|}
\end{split}
\end{equation}
in which the exponents are
\begin{equation}\label{eq:DPABdGEexponents}
\begin{split}
\alpha_{j \pm} & = \sqrt{\hat{k}_j^2 + 1 \pm \sqrt{1+\hat{\omega}_j^2} } \\
\beta_{j \pm} & = \sqrt{\hat{k}_j^2 +\beta^2 \pm \hat{\omega}_j} \\
\end{split},
\end{equation}
Note that the functions $\Sigma_{j'k}$ and $\Delta_{j'k}$ need no superscript because no approximation is made in their differential equations. 

We note that the following relations hold among the coefficients (since exponentials with different decay are, in general, independent), 
\begin{equation}\label{eq:DPABdGESigmaDeltacoefficients}
\begin{split}
B_{jk \pm} & = \frac{1 \mp \sqrt{1+\hat{\omega}_j^2}}{ \hat{\omega}_j} A_{jk \pm} \equiv \pi_{j\mp}A_{jk \pm}\\
D_{jk \pm} & = \mp C_{jk \pm}  
\end{split}.
\end{equation}
There remain 4 undetermined coefficients (the $A$ and the $C$) for each BEC component.

It is instructive to give also the zeroth-order solutions in terms of the $u_{jk}$ and $v_{jk}$, because some of these are just single exponentials,
\begin{equation}\label{eq: DPABdGEuvsolution}
\begin{split}
u^{(0)}_{jk} &= a_{jk+} e^{-  \alpha_{j+} \left| \rho_j \right|} + a_{jk-} e^{-  \alpha_{j-}\left| \rho_j \right|}  \\
v^{(0)}_{jk} &= b_{jk+} e^{-   \alpha_{j+} \left| \rho_j \right|} + b_{jk-} e^{-  \alpha_{j-} \left| \rho_j \right|}  \\
u_{j'k} &= C_{j'k-} e^{-  \beta_{j'-} \left| \rho_{j'} \right|} \\
v_{j'k} &= C_{j'k+} e^{-  \beta_{j'+} \left| \rho_{j'} \right|} 
\end{split}
\end{equation}
where the $C$ are the same as in \eqref{eq:DPABdGESigmaDeltasolution}
and
\begin{equation}\label{eq:DPABdGEexpoentsandcoefficients}
b_{jk \pm}  = \left(\hat{\omega}_j \pm \sqrt{1+\hat{\omega}_j^2} \right) a_{jk \pm} 
\end{equation}

Now we are prepared to return to the problem of solving the first two DPA BdG equations in \eqref{eq:DPABdGESigmaDelta} considering the term $2(\phi_{j0} -1)\Delta_{jk}$ perturbatively. Note that this term is small in the weak-segregation limit, where $\beta$ is small, but we are interested in using the approximation for all $\beta$. Therefore we assume $\epsilon_{j0} = \phi_{j0} -1$ to be small and perform a perturbation calculation to first order in the amplitude of this quantity, which we call $q$, i.e.,
\begin{equation}
q \equiv \frac{\beta}{\sqrt{2}+ \beta}
\end{equation} 
We therefore expand,
\begin{equation}\label{eq:DPABdGFirstOrdersolutionPrelude}
\begin{split}
\Sigma_{jk}& =\Sigma^{(0)}_{jk} + q\Sigma^{(1)}_{jk} \\
\Delta_{jk}& = \Delta^{(0)}_{jk} + q\Delta^{(1)}_{jk}  
\end{split}
\end{equation}

Our main task now is to find the solutions of the {\em inhomogeneous} system
\begin{equation}\label{eq:DPABdGESigmaDeltaInhom}
\begin{split}
\left( \mathcal{A}_{jk} + 2\right) \Sigma^{(1)}_{jk} &= \hat{\omega}_j \Delta^{(1)}_{jk} \\
 \mathcal{A}_{jk}  \Delta^{(1)}_{jk} &=   \hat{\omega}_j \Sigma^{(1)}_{jk} +2e^{- \sqrt{2}\, \left| \rho_j \right|}\;\Delta^{(0)}_{jk}
\end{split}
\end{equation}
Note that we have used the explicit form \eqref{eq:stationarysolutionsDPAGPE} of $\phi_{j0}$ in DPA and $\Delta^{(0)}_{jk}$ is given by the second equation of \eqref{eq:DPABdGESigmaDeltasolution}.
The solutions of these ``first-order" equations consist of a particular solution complemented by an arbitrary solution of the {\em homogeneous} system, which is by definition the system \eqref{eq:DPABdGESigmaDeltaInhom} but without the last term. 

Summarizing, we can write the general solution to the first two inhomogeneous DPA BdG equations in \eqref{eq:DPABdGESigmaDelta} formally as follows, to first order in $q$,
\begin{equation}\label{eq:DPABdGFirstOrdersolution}
\begin{split}
\Sigma_{jk}& =\Sigma^{(0)}_{jk} + q\Sigma^{(1)}_{jk} = \Sigma^{(0)}_{jk} + q(\Sigma^{(p)}_{jk} + \Sigma^{(h)}_{jk})\\
\Delta_{jk}& = \Delta^{(0)}_{jk} + q\Delta^{(1)}_{jk} = \Delta^{(0)}_{jk} + q(\Delta^{(p)}_{jk}  + \Delta^{(h)}_{jk})
\end{split}
\end{equation}
The superscript $p$ refers to the particular solution of the  system \eqref{eq:DPABdGESigmaDeltaInhom} and the superscript $h$ refers to an arbitrary solution of the homogeneous system, which coincides formally with an arbitrary solution of the zeroth-order system \eqref{eq:DPABdGESigmaDeltasolution}. 
After inspection of the system \eqref{eq:DPABdGESigmaDeltaInhom} we can write the particular solutions in the form
\begin{equation}\label{eq:DPABdGFirstOrderparticular}
\begin{split}
\Sigma^{(p)}_{jk}& = \left (S_{jk+} e^{- \alpha_{j+} \left| \rho_j \right|}+ S_{jk-} e^{- \alpha_{j-} \left| \rho_j \right|}  \right )e^{- \sqrt{2}\, \left| \rho_j \right|} \\
\Delta^{(p)}_{jk}& = \left ( T_{jk+} e^{-\alpha_{j+} \left| \rho_j \right|}+ T_{jk-} e^{-  \alpha_{j-} \left| \rho_j \right|}  \right )e^{- \sqrt{2}\, \left| \rho_j \right|},
\end{split}
\end{equation}
where the (finite) coefficients $S$ and $T$ are found to satisfy the following equations
\begin{eqnarray}
S_{jk\pm} (\hat{k}_j^2 - (\sqrt{2}\, +  \alpha_{j\pm} )^2+2)&= &\hat{\omega}_jT_{jk\pm}\label{particularequations1}\\
T_{jk\pm}(\hat{k}_j^2 - (\sqrt{2}\, +  \alpha_{j\pm} )^2 )& = &\hat{\omega}_j S_{jk\pm} + 2 B_{jk\pm}\label{particularequations2}
\end{eqnarray}
This leads to the following expressions, which are useful for $\hat{k}_j \neq   0
\neq \hat{\omega}_j$,
\begin{eqnarray}\label{particularCoeff}
S_{jk\pm} &= &\frac{2\;\hat{\omega}_j\; B_{jk\pm}}{(\hat{k}_j^2 - (\sqrt{2}\, +  \alpha_{j\pm} )^2 )(\hat{k}_j^2 - (\sqrt{2}\, +  \alpha_{j\pm} )^2+2) - \hat{\omega}_j^2}\\
T_{jk\pm} & = &\frac{2 \left (\hat{k}_j^2 - (\sqrt{2}\, +  \alpha_{j\pm} )^2+2\right )B_{jk\pm}}{(\hat{k}_j^2 - (\sqrt{2}\, +  \alpha_{j\pm} )^2 )(\hat{k}_j^2 - (\sqrt{2}\, +  \alpha_{j\pm} )^2+2) - \hat{\omega}_j^2}
\end{eqnarray}

The arbitrary homogeneous solutions present in the ``first-order correction" are of the form
\begin{equation}\label{eq:DPABdGESigmaDeltasolutionArbitrary}
\begin{split}
\Sigma^{(h)}_{jk} &= A^{(h)}_{jk+} e^{-  \alpha_{j+} \left| \rho_j \right|} + A^{(h)}_{jk-} e^{- \alpha_{j-} \left| \rho_j \right|}  \\
\Delta^{(h)}_{jk} &= B^{(h)}_{jk+} e^{-   \alpha_{j+} \left| \rho_j \right|} + B^{(h)}_{jk-} e^{-  \alpha_{j-} \left| \rho_j \right|}  
\end{split}
\end{equation}
Note that without loss of generality, we can choose some of the coefficients to be equal to zero. In particular, we have the freedom to choose one of the coefficients in 
$\Sigma_{jk} $ or $\Delta_{jk} $ to set the scale of the excitation. If we choose to fix $A_{jk+}$, we do not need to introduce any first-order correction to it and can thus set $A^{(h)}_{jk+}=0$. This entails also $B^{(h)}_{jk+} =0$ in view of \eqref{eq:DPABdGESigmaDeltacoefficients}. Since, furthermore, $A^{(h)}_{jk-}$ is related to $B^{(h)}_{jk-} $ through the same constraint \eqref{eq:DPABdGESigmaDeltacoefficients}, we observe that the presence of the homogeneous solutions generates only one additional unknown, say $B^{(h)}_{jk-} $.

We now proceed to the discussion of the full solution of our system of 4 differential equations. Even though the two pairs of equations are independent, their solutions get coupled through suitable boundary conditions, which reduce the number of undetermined coefficients.  Indeed, we must invoke several continuity conditions at the interface. Firstly, for $K < \infty$ (excluding the case of total segregation) it is reasonable to impose {\em continuity} of all solutions at the interface, since the zero modes are continuous at the interface, in GP theory as well as in the DPA, and we aim at building a dynamical DPA with no more singularities than those present in the static DPA.
This leads to the following relations,
\begin{eqnarray}
A_{jk+} + A_{jk-} + q( S_{jk+} + S_{jk-} + A^{(h)}_{jk-})&=& C_{jk+} + C_{jk-} ,  \label{continuityrelations1}\\
B_{jk+} + B_{jk-} + q (T_{jk+} + T_{jk-} + B^{(h)}_{jk-} )&=& -C_{jk+} + C_{jk-},\label{continuityrelations2}
\end{eqnarray}
while the coefficients $A$ and $B$ of the ``zeroth-order" homogeneous solutions satisfy \eqref{eq:DPABdGESigmaDeltacoefficients}. 

Next, we require continuity of the ``hydrodynamic" amplitude, i.e., the first derivative of $\Delta_{jk}$, at the interface. This gives
\begin{equation}\label{continuityderivative}
\alpha_{j+} B_{jk+}  + \alpha_{j-}  B_{jk-} + q ((\sqrt{2}\, + \alpha_{j+})T_{jk+}  + (\sqrt{2}\, + \alpha_{j-})T_{jk-} + \alpha_{j-}B^{(h)}_{jk-}) = \beta_{j+}C_{jk+}  - \beta_{j-} C_{jk-}
\end{equation}
Now we are left with (only) one undetermined coefficient per BEC component, assuming that we fix $A_{jk+}$.

Upon inspection of our system of differential equations with these 3 boundary conditions, we conclude that it is not necessary to include an extra homogeneous solution in the first-order perturbation scheme. The boundary conditions can all be satisfied using just the particular solution as the perturbation. So we can simplify the analysis and choose $B^{(h)}_{jk-}=0$, which implies also $A^{(h)}_{jk-}=0$.

We are now in a position to generate all the modes analytically, within the DPA, for values of the reduced wave numbers  $\hat{k}_j$ and reduced frequencies $\hat{\omega}_j$ that are arbitrarily chosen, as long as the exponents $\alpha_{j \pm}$ and $\beta_{j \pm}$, defined in \eqref{eq:DPABdGEexponents}, are real numbers. Clearly, this mathematical set of modes is larger than the physical subset that satisfies the yet undetermined dispersion relation (the physical constraint expressing the dependence of the frequency on the wave number). We will turn to the derivation of the dispersion relation later, but for now, we consider the frequencies and wave numbers as independent parameters and present figures illustrating functions $\Sigma_{jk}$ and $\Delta_{jk}$ that satisfy all our approximations so far. The functions are of the form \eqref{eq:DPABdGFirstOrdersolution}, with 8 coefficients that must satisfy the 7 equations \eqref{eq:DPABdGESigmaDeltacoefficients}, \eqref{particularCoeff}, \eqref{continuityrelations1},  \eqref{continuityrelations2} and \eqref{continuityderivative}. There remains one undetermined coefficient that, without loss of generality, can be attributed an arbitrary value, which sets the scale for the mode.

Fig.1 shows arbitrary examples of $\Sigma_{1k}$-modes for the symmetric case $\xi_1 = \xi_2 \equiv \xi$, implying $\rho_1 = \rho_2 \equiv \rho$. For this case the modes satisfy reflection antisymmetry $\Sigma_{2k} (\rho) = -\Sigma_{1k}(-\rho)$.  For the wave number the choice $\hat{k}_1=\hat{k}_2\equiv k\xi = 1$ has been made. Furthermore, the chosen reduced frequencies are $\hat{\omega}_1= \hat{\omega}_2 = 1$. The freedom to choose the coefficient $A_{jk+}$ has been exploited by fixing it to $\sqrt{2}\beta/(\sqrt{2}+ \beta)$, with $\beta = \sqrt{K-1}$, in accord with the amplitude implied by \eqref{generalformDS}. Four segregation strengths are presented: weak segregation ($K=1.1$), intermediate segregation ($K=1.5$ and $K=3$) and strong segregation ($K=10$). Note that the curve for $K=3$ is nearly, but not fully, symmetric about the interface. The reason for this (near-)symmetry is that, for $K=3$, the amplitudes of the static DPA order parameters are equal to $1/2$ at the interface since $\beta= \sqrt{2}$. 

\begin{figure}
\centering
\includegraphics[width=0.77\textwidth]{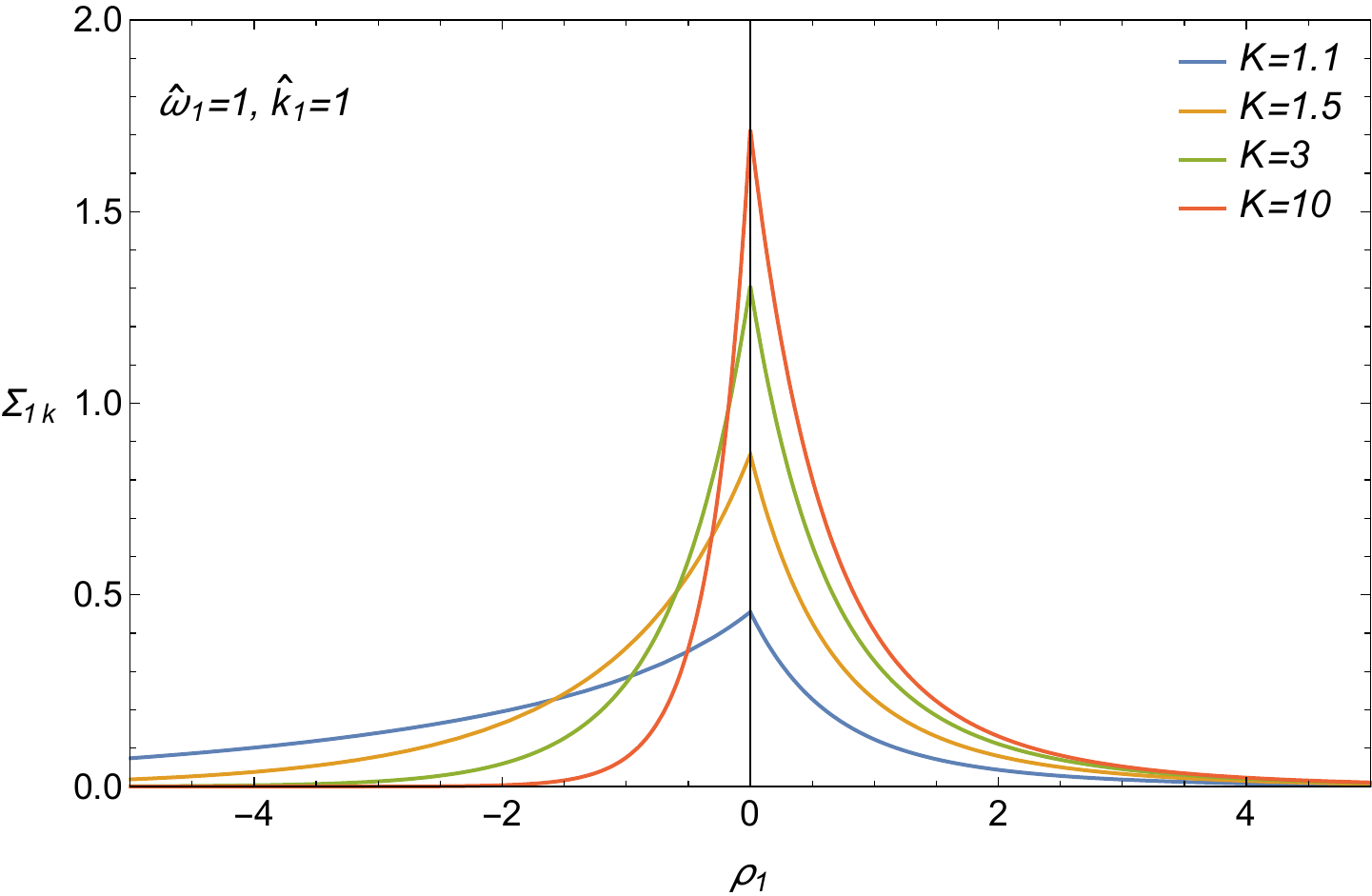}
\caption{Spatial dependence of the $\Sigma_{1k}$-mode for various condensate segregation strengths. Calculations are shown for the symmetric case $\xi_1=\xi_2\equiv \xi$. The curve with the lowest peak corresponds to reduced interaction strength $K=1.1$ (weak segregation), the two intermediate peaks to $K=1.5$ and $K=3$ and the highest peak to $K=10$ (strong segregation). Note that $\Sigma_{1k}$ is continuous, but its first derivative is discontinuous, at the interface, which is located at $\rho_1=0$.}
\label{fig:1}
\end{figure}

For the same choices of parameters, Fig.2 shows examples of $\Delta_{1k}$-modes. For the symmetric case $\xi_1 = \xi_2 \equiv \xi$ the modes satisfy reflection antisymmetry $\Delta_{2k} (\rho) = -\Delta_{1k}(-\rho)$.  Figs.3 and 4 show examples of $\Sigma_{1k}$-modes and $\Delta_{1k}$-modes, respectively, at constant interaction strength $K$ and at constant reduced wave number, but with varying reduced frequency.

\begin{figure}
\centering
\includegraphics[width=0.77\textwidth]{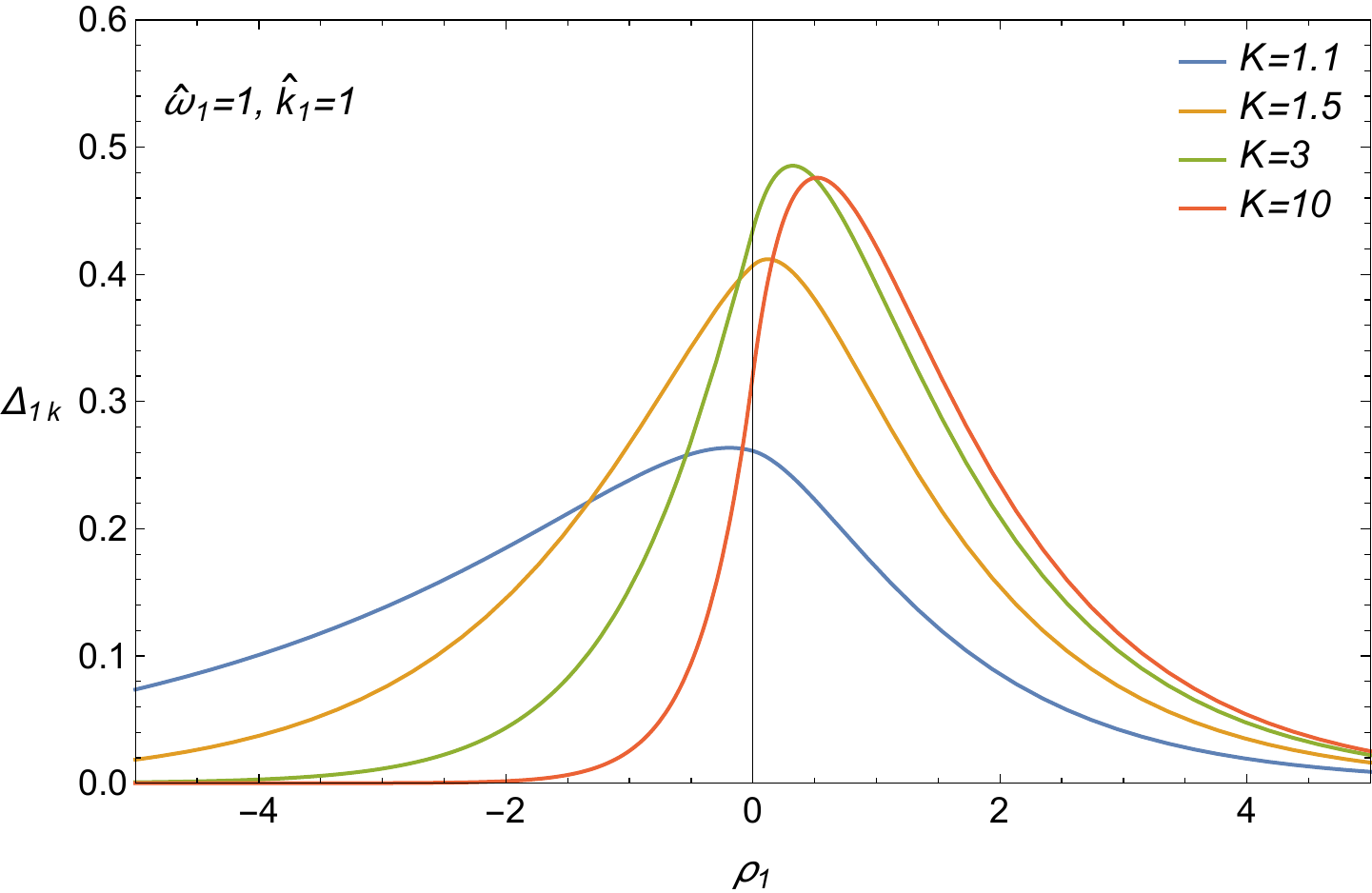}
\caption{Spatial dependence of the $\Delta_{1k}$-mode for various condensate segregation strengths. The curve with the mildest slope at the interface corresponds to reduced interaction strength $K=1.1$ (weak segregation), the intermediate curves to $K=1.5$ and $K=3$ and the curve with the steepest slope at the interface to $K=10$ (strong segregation). Note that $\Delta_{1k}$ and its first derivative are continuous at the interface, which is located at $\rho_1=0$.}
\label{fig:2}
\end{figure}

\begin{figure}
\centering
\includegraphics[width=0.77\textwidth]{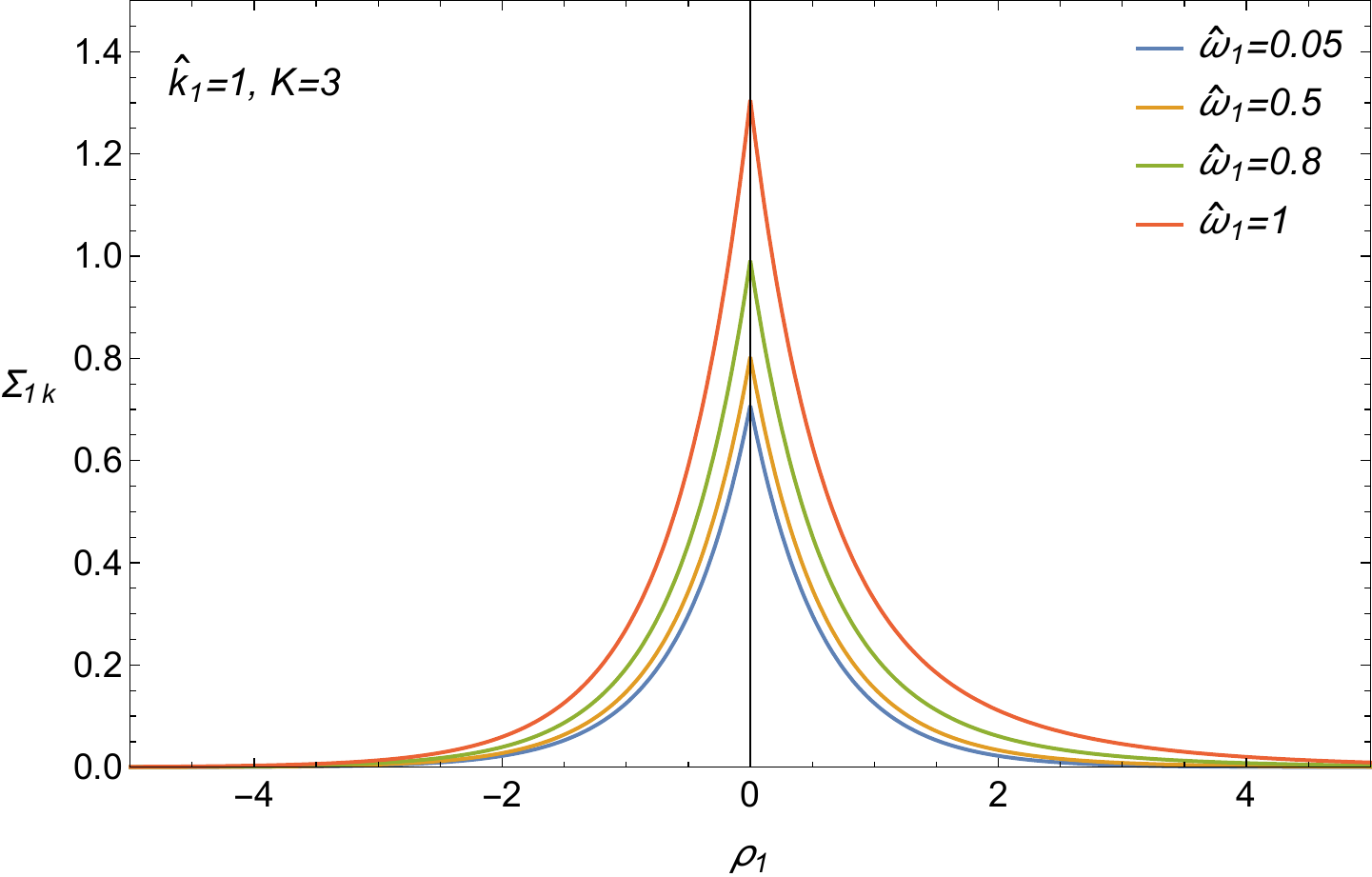}
\caption{Spatial dependence of the $\Sigma_{1k}$-mode for various reduced excitation frequencies and for $K=3$.  The lower curve  corresponds to the lowest frequency $\hat{\omega}_1= 0.05$, the intermediate curves to $\hat{\omega}_1 =0.5$ and $\hat{\omega}_1=0.8$ and the upper curve  to $\hat{\omega}_1=1$. Note that the curves are nearly, but not fully, symmetric about the interface.}
\label{fig:3}
\end{figure}

\begin{figure}
\centering
\includegraphics[width=0.77\textwidth]{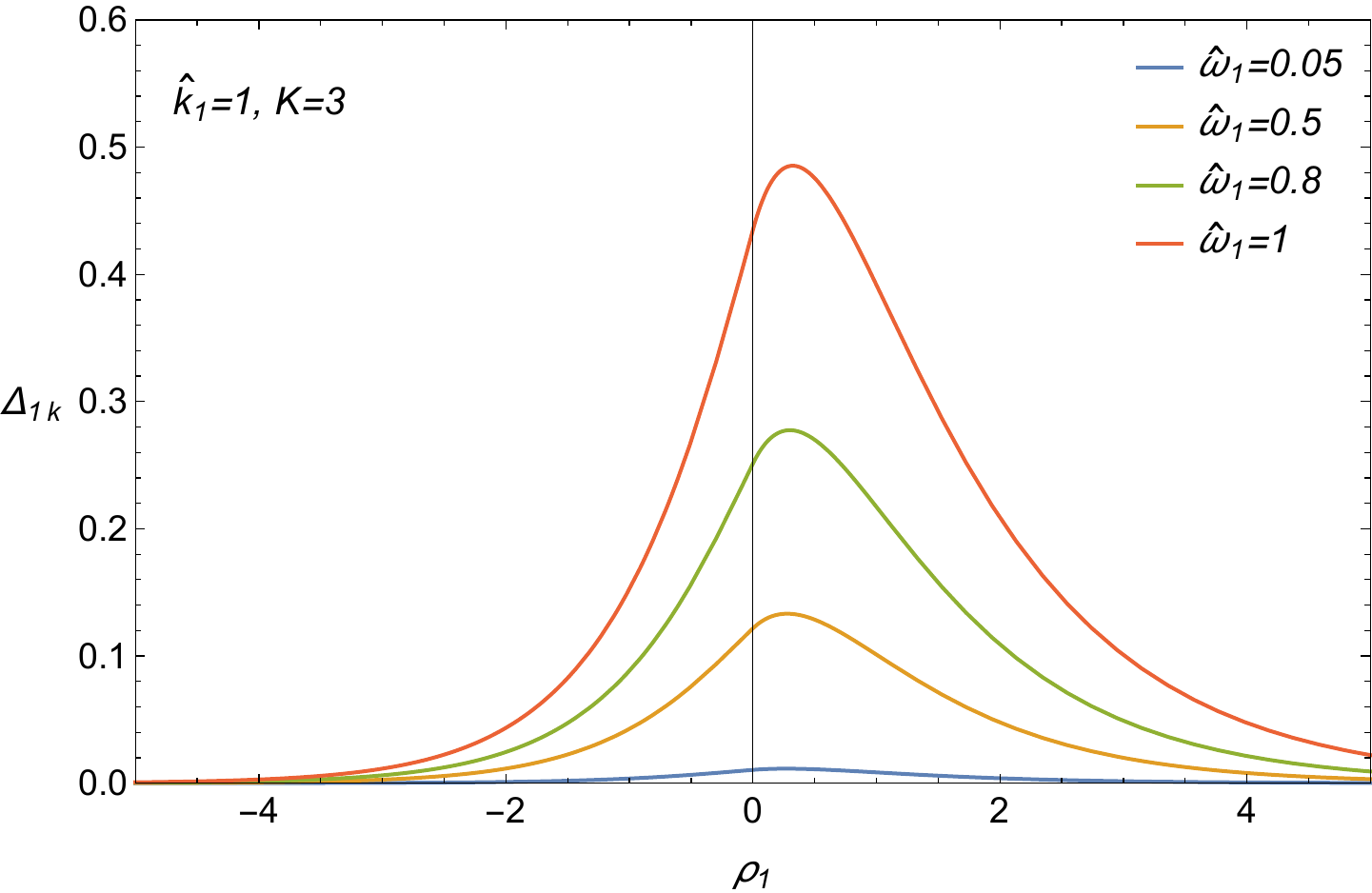}
\caption{Spatial dependence of the $\Delta_{1k}$-mode  for various reduced excitation frequencies  and for $K=3$. The lower curve  corresponds to the lowest frequency $\hat{\omega}_1= 0.05$, the intermediate curves to $\hat{\omega}_1 =0.5$ and $\hat{\omega}_1=0.8$ and the upper curve  to $\hat{\omega}_1=1$.}
\label{fig:4}
\end{figure}

\subsection{Physical DPA solutions for ripplons in the long-wavelength limit}
We now address the long-wavelength limit. We assume that when $k$ tends to zero, also $\omega$ tends to zero, anticipating the existence of a dispersion relation. In fact, as is appropriate for ripplons, we assume that $\omega$ tends to zero more rapidly than linearly in $k$, and we eventually attempt to capture the correct exponent from our calculations.

To make progress we expand all the auxiliary quantities in $\hat{\omega}$ and $\hat{k}$ to the lowest relevant orders, suppressing momentarily  the subscript $j$,
\begin{eqnarray}
\pi_- &=& - \frac{\hat{\omega}}{2} + {\cal O}(\hat{\omega} ^3) \label{useful1}  \\
\pi_+& =& \frac{2}{\hat{\omega}} +  \frac{\hat{\omega}}{2}   + {\cal O}(\hat{\omega} ^3) \label{useful2}  \\
\alpha_+ &=& \sqrt{2}\, \left (1+  \frac{\hat{k}^2}{4} + \frac{\hat{\omega}^2}{8} + {\cal O}(\hat{k}^4, \hat{k}^2 \hat{\omega}^2, \hat{\omega}^4) \right )\label{useful3} \\
\alpha_- &=& \hat{k} -  \frac{\hat{\omega}^2}{4\hat{k}} + {\cal O}\left (\frac{\hat{\omega}^4}{\hat{k}^3}\right )\label{useful4} \\
\beta_{\pm} &=& \beta \left ( 1 + \frac{1}{2\beta^2} (\hat{k}^2 \pm \hat{\omega}) + {\cal O}(\hat{\omega}^2)\right )\label{useful5} 
\end{eqnarray}

These expressions allow us to check explicitly the zero modes within our DPA strategy. Considering the limit $\hat{k}_j \rightarrow 0$ and $\hat{\omega}_j \rightarrow 0$, from \eqref{particularequations2} we readily obtain
\begin{equation}\label{first}
T_{j0-} =  - B_{j0-},
\end{equation} 
with $B_{j0-}$ finite (or zero). Furthermore, $B_{j0+}$ must vanish, in view of \eqref{eq:DPABdGESigmaDeltacoefficients} and \eqref{useful1}, and the fact that $A_{j0+}$ is finite (or zero). So,
\begin{equation}
B_{jk+}  \rightarrow B_{j0+} = 0, \;  \mbox{for} \;\hat{k}_j, \hat{\omega}_j \rightarrow 0
\end{equation}
Consequently, using also \eqref{particularequations2} we obtain
\begin{equation}\label{third}
T_{j0+} =  0
\end{equation} 
Furthermore, regardless of the (finite or zero) value of $T_{j0+} $, \eqref{particularequations1} implies
\begin{equation}\label{fourth}
S_{j0+} =  0
\end{equation} 
However, the determination of $S_{j0-}$ is not straightforward and requires a careful examination of the zero-mode limit. Using \eqref{particularCoeff} we obtain, using the symbol ``$\sim$" for ``asymptotically equal to" (whereas ``$\propto$" is used for ``proportional to"),
\begin{equation}\label{fifth}
S_{jk-} \sim    \frac{\hat{\omega}_j}{2\sqrt{2}\,\,\hat{k}_j}B_{jk-}, \;  \mbox{for} \;\hat{k}_j, \hat{\omega}_j \rightarrow 0
\end{equation} 
which tends to zero, regardless of the (finite or zero) value of $B_{j0-}$, provided $\hat{\omega}_j/\hat{k}_j \rightarrow 0$ in the long wavelength limit. 

The information gathered so far already narrows down considerably the form of the zero modes. We get
\begin{equation}\label{zeromodeformS}
\Sigma_{j0} = \Sigma^{(0)}_{j0} + q\Sigma^{(p)}_{j0}  = A_{j0+} e^{- \sqrt{2}\, \left| \rho_j \right|} + A_{j0-},
\end{equation}
and
\begin{equation}\label{zeromodeformD}
\Delta_{j0} = \Delta^{(0)}_{j0} + q\Delta^{(p)}_{j0} =  B_{j0-}(1 +  \epsilon_{j0})  
\end{equation}
Since $\Delta_{j0}$ must be proportional to the DPA order parameter and $\Sigma_{j0} $ to its derivative, owing to \eqref{DPAzeroD} and \eqref{DPAzeroS}, respectively, we must ask 
\begin{equation}\label{eight} 
A_{j0-} =0,
\end{equation} 
and, moreover, $A_{jk-}$ must vanish at least as fast as $\hat{\omega}_j $, for  $\hat{\omega}_j \rightarrow 0$, in order to avoid a divergence of $B_{j0-}$, in view of 
\eqref{eq:DPABdGESigmaDeltacoefficients}.

Next, \eqref{continuityrelations1} implies
\begin{equation}\label{sixth}
C_{j0+} + C_{j0-} =   A_{j0+},
\end{equation}
and \eqref{continuityrelations2} leads to
\begin{equation}\label{seventh}
-C_{j0+}  + C_{j0-} = \frac{\sqrt{2}\,}{\sqrt{2}\, + \beta}  \; B_{j0-}
\end{equation}

The foregoing results allow us to retrieve the correct zero modes within the DPA,
\begin{equation}\label{zeromodefinal}
\begin{split}
\Sigma_{j0}& =  A_{j0+} e^{- \sqrt{2}\, \left| \rho_j \right|}\\
\Delta_{j0} & =  B_{j0-}\left (1 -  \frac{\beta}{\sqrt{2}\, + \beta}\, e^{- \sqrt{2}\, \left| \rho_j \right|}\right ) \\
\Sigma_{j'0} &= A_{j'0+} e^{-  \beta  \left|\rho_{j'} \right|}\\
\Delta_{j'0} &= 
\frac{\sqrt{2}\,}{\sqrt{2}\, + \beta}  B_{j'0-}\,e^{-  \beta  \left|\rho_{j'} \right|}
\end{split}
\end{equation}
Note that some of the amplitudes can be zero depending on which kind of zero modes are excited. For a uniform translation along $z$, $A_{j0+} \neq 0 = B_{j0-}$, and for a uniform phase rotation $A_{j0+} =  0 \neq B_{j0-}$. 

The coefficients that vanish in the zero-mode limit will henceforth be regarded as dynamical amplitudes, and the non-vanishing coefficients will be regarded as static amplitudes. We now combine the relations among the coefficients, \eqref{continuityrelations1} and \eqref{continuityrelations2}, with the expansions \eqref{useful1}-\eqref{useful5} to lowest orders in $\hat{\omega}$ and $\hat{k}$ and substitute them in 
\eqref{continuityderivative}. With the help of the auxiliary expansions
\begin{eqnarray}
T_{jk+} &= &\frac{\hat{\omega}_j}{8} A_{jk+}\left (1+ {\cal O}(\hat{k}_j^2,  \hat{\omega}_j^2) \right )\\
T_{jk-} & = & B_{jk-} \left (-1+ \sqrt{2}\, \hat{k}_j + {\cal O}\left (\frac{\hat{\omega}_j^2}{\hat{k}_j},\frac{\hat{\omega}_j^4}{\hat{k}_j^4}\right )\right ),  
\end{eqnarray}
this leads to 
\begin{equation}\label{eight}
(1+q(1+\sqrt{2}\;\beta))B_{jk-} = \left (\frac{1}{2\beta} + \frac{\sqrt{2}}{2}+ \frac{\beta}{2} + q(\frac{\beta}{8} - \frac{\sqrt{2}}{4} )\right )  \frac{\hat{\omega}_j}{\hat{k}_j} A_{jk+} \; + \; {\cal O}(\hat{\omega}_j)
\end{equation} 
We now note that $\beta$ can be expanded in $q$ as follows
\begin{equation}\label{betaxp}
\beta \equiv \frac{\sqrt{2}\;q}{1-q} = \sqrt{2}\; q (1+ q + {\cal O}(q^2))
\end{equation}
Applying this in \eqref{eight}  leads to
\begin{equation}\label{eightbis}
B_{jk-} \equiv {\cal J}(q) \,  \frac{\hat{\omega}_j}{\hat{k}_j} A_{jk+} \; + \; {\cal O}(\hat{\omega}_j),
\end{equation} 
with 
\begin{equation}\label{Jdef}
{\cal J}(q)^{-1} =  2\sqrt{2}\,q +  {\cal O}(q^{3}) 
\end{equation}
Since the next-to-leading term in this expression is of order $q^3$, we obtain a good approximation by keeping only the leading term. This leads to the important result
\begin{equation}\label{Jdefactual}
J(\beta)^{-1} \equiv {\cal J}(q)^{-1}  \approx  \frac{2\sqrt{2}\,\beta}{\sqrt{2} + \beta}  
\end{equation}

We conclude that also $B_{jk-}$ is a dynamical amplitude, since $B_{j0-}$ vanishes, provided $\hat{\omega}_j/\hat{k}_j \rightarrow 0$ in the long wavelength limit. In hindsight, this is not surprising because our excitation does not contain a uniform phase rotation in the zero-mode limit, but only a translational mode with non-zero (fixed) amplitude $\zeta_0$ and diverging wavelength. 

In sum, the zero modes \eqref{zeromodefinal} for our specific ripplon problem reduce to the simple forms
\begin{equation}\label{zeromodefinalspecific}
\begin{split}
\Sigma_{j0}& =  i \varrho_{j0}^{(R)}  \frac{(-1)^{j+1}\sqrt{2}\,\beta}{\sqrt{2}\,+\beta} e^{- \sqrt{2}\, \left| \rho_j \right|}\\
\Delta_{j0} & =  0 \\
\Sigma_{j'0} &= i \varrho_{j'0}^{(R)} \frac{(-1)^{j'+1}\sqrt{2}\,\beta}{\sqrt{2}\,+\beta} e^{-  \beta  \left|\rho_{j'} \right|}\\
\Delta_{j'0} &= 0,
\end{split}
\end{equation}
where we have exploited the freedom to fix the static amplitude $A_{j0+}$ with the help of the general form \eqref{generalformDS} for the case $F=0$, i.e., in the absence of structure fluctuations.

The relations among coefficients derived so far allow one to express all the coefficients in terms of $A_{jk+}$, whose static limit $A_{j0+}$ is known. 
We thus obtain to first order in $q$, using ${\cal J}(q) = (2\sqrt{2} q)^{-1}(1 + {\cal O}(q^2))$, the following explicit forms for the long-wavelength limit of all the relevant functions contained in \eqref{eq:DPABdGFirstOrdersolution},
\begin{equation}\label{DPABdGFirstOrderSolutions}
\begin{split}
\Sigma^{(0)}_{jk} & \propto i \varrho_{j0}^{(R)} (-1)^{j+1} \left ( \frac{\sqrt{2}\,\beta}{\sqrt{2}+\beta}  e^{- \alpha_{j+}  \left| \rho_j \right|} +  \frac{\hat{\omega}_j^2}{4\hat{k}_j}e^{- \alpha_{j-} \left| \rho_j \right|}\right ) + ...\\
\Sigma^{(p)}_{jk} & \propto i \varrho_{j0}^{(R)}  (-1)^{j+1}\left (-\frac{\sqrt{2}\,\beta}{\sqrt{2}+\beta}  \frac{\hat{\omega}_j^2}{48} e^{- (\alpha_{j+}+\sqrt{2}) \left| \rho_j \right|} + \frac{\sqrt{2}}{8} \,\frac{\hat{\omega}_j^2}{\hat{k}_j^2} e^{-  (\alpha_{j-} +\sqrt{2}) \left| \rho_j \right|}\right ) + ...\\
\Delta^{(0)}_{jk} &\propto i \varrho_{j0}^{(R)}  (-1)^{j+1}\left (-\frac{\sqrt{2}\,\beta}{\sqrt{2}+\beta}  \frac{\hat{\omega}_j}{2} e^{-  \alpha_{j+} \left| \rho_j \right|} +   \frac{\hat{\omega}_j}{2\hat{k}_j} e^{- \alpha_{j-} \left| \rho_j \right|}\right ) + ...\\
\Delta^{(p)}_{jk}&\propto i \varrho_{j0}^{(R)} (-1)^{j+1}\left (\frac{\sqrt{2}\,\beta}{\sqrt{2}+\beta}  \frac{\hat{\omega}_j}{8} e^{-  (\alpha_{j+}+\sqrt{2}) \left| \rho_j \right|} -  \frac{\hat{\omega}_j}{2\hat{k}_j} e^{-  (\alpha_{j-}+\sqrt{2}) \left| \rho_j \right|}\right )  + ...\\
\Sigma_{j'k} &\propto  i \varrho_{j'0}^{(R)}\frac{(-1)^{j'+1} \sqrt{2}\,\beta}{2(\sqrt{2}+\beta)} \left ((1- \frac{\hat{\omega}_{j'}}{2\beta\hat{k}_{j'}} + \frac{\hat{\omega}_{j'}^2}{8\hat{k}_{j'}^2}  ) e^{-\beta_{j'+}  \left|\rho_{j'} \right|}  
  +  (1+\frac{\hat{\omega}_{j'}}{2\beta\hat{k}_{j'}} + \frac{\hat{\omega}_{j'}^2}{8\hat{k}_{j'}^2} ) e^{-\beta_{j'-}  \left|\rho_{j'} \right|} \right ) + ... \\
\Delta_{j'k} &\propto  i \varrho_{j'0}^{(R)}\frac{ (-1)^{j'+1}\sqrt{2}\,\beta}{2(\sqrt{2}+\beta)} \left ((-1+ \frac{\hat{\omega}_{j'}}{2\beta\hat{k}_{j'}} - \frac{\hat{\omega}_{j'}^2}{8\hat{k}_{j'}^2}) e^{-\beta_{j'+}  \left|\rho_{j'} \right|}  
 +(1+\frac{\hat{\omega}_{j'}}{2\beta\hat{k}_{j'}}  + \frac{\hat{\omega}_{j'}^2}{8\hat{k}_{j'}^2} ) e^{-\beta_{j'-}  \left|\rho_{j'} \right|} \right ) + ..., 
\end{split}
\end{equation}
where the overall proportionality factor is an unimportant unknown function of $\hat{\omega}_j$ and $\hat{k}_j $ which must approach unity in the long wavelength limit.    

Next, limiting ourselves to the leading terms, we obtain the  {\em asymptotic forms}, for $\hat{k}_j, \hat{\omega}_j \rightarrow 0$, in a form that allows one to identify the DPA versions of the functions $f$ and $F$, defined through \eqref{etaandtheta}, that pertain to the phase fluctuations and shape fluctuations, respectively, of the dynamically perturbed interface. We obtain, after summation according to \eqref{eq:DPABdGFirstOrdersolution}, and recalling $\Sigma^{(h)}_{jk} = \Delta^{(h)}_{jk}=0$,
\begin{equation}\label{DPABdGFirstOrderSolution}
\begin{split}
\Sigma_{jk}& \sim i \varrho_{j0}^{(R)}  \frac{(-1)^{j+1}\sqrt{2}\,\beta}{\sqrt{2}+\beta} e^{- \sqrt{2}\, \left| \rho_j \right|} 
\left (1+ \frac{\hat{\omega}_j^2}{8\hat{k}_j^2} e^{- \hat{k}_j \left| \rho_j \right|}\right ) \\
&\approx  i \varrho_{j0}^{(R)} \partial_{\rho_j} \phi_{j0}\left (1+  \frac{\hat{\omega}_j^2}{8\hat{k}_j^2} \right ) \\
\Delta_{jk}& \sim i \varrho_{j0}^{(R)} (-1)^{j+1} \frac{\hat{\omega}_j}{2\hat{k}_j} e^{- \hat{k}_j \left| \rho_j \right|}(1 - \frac{\beta}{\sqrt{2}+\beta}e^{- \sqrt{2}\, \left| \rho_j \right|})\\
&= i \varrho_{j0}^{(R)}\phi_{j0}(-1)^{j+1} \frac{\hat{\omega}_j}{2\hat{k}_j} e^{- \hat{k}_j \left| \rho_j \right|} \\
\Sigma_{j'k}& \sim  i \varrho_{j'0}^{(R)} \frac{(-1)^{j'+1}\sqrt{2}\,\beta}{\sqrt{2}+\beta} e^{- \beta \left| \rho_{j'} \right|} \left (1+\frac{\hat{\omega}_j'^2}{8\hat{k}_j'^2} \right )\\
&= i \varrho_{j'0}^{(R)} \partial_{\rho_j'} \phi_{j'0} \left (1+  \frac{\hat{\omega}_j'^2}{8\hat{k}_j'^2} \right )\\
\Delta_{j'k} &  
\sim i \varrho_{j'0}^{(R)} \frac{(-1)^{j'+1} \sqrt{2}\,\beta}{\sqrt{2}+\beta} e^{- \beta \left| \rho_{j'} \right|} \left (   \frac{\hat{\omega}_{j'}}{2\beta\,\hat{k}_{j'}}    +  \sinh (\frac{\hat{\omega}_{j'}}{2\beta}  \left| \rho_{j'} \right|)  \right )   \\
&\approx   i \varrho_{j'0}^{(R)} \phi_{j'0} (-1)^{j'+1}\left (  \frac{\hat{\omega}_{j'}}{2\hat{k}_{j'}} + \frac{\hat{\omega}_{j'}}{2}  \left| \rho_{j'} \right|)\right )
\end{split}
\end{equation}
Note that the first term in the first right-hand-side in $\Delta_{jk}$ is the dominant one in the bulk region far from the interface. This term  expresses the slow decay (decay length $\approx $ wavelength) of the $\Delta$ modes. This is exactly what one should expect in view of a qualitative analysis (not performed here) of the ripplon excitation in GP BdG theory (beyond DPA) in the distant field, far away from the interface.  Note also the important last term in the right-hand-side in  $\Delta_{j'k}$. It gives rise to a position dependence of the function $f_{j'k}$, essential for obtaining a continuous superfluid velocity across the interface. 

From these expressions we can extract the leading term(s) of the functions $F_{jk}$ and $f_{jk}$ within the DPA, using the correspondence \eqref{generalformDS}. They take the following form 
\begin{equation}\label{DPAofFandf}
\begin{split}
F_{jk}^{(DPA)}& =  
 \frac{\hat{\omega}_j^2}{8\hat{k}_j^2} \partial_{\rho_j} \ln \phi_{j0}\\
f_{jk}^{(DPA)}& = (-1)^{j+1}\frac{\hat{\omega}_j}{2\hat{k}_j} e^{- \hat{k}_j \left| \rho_j \right|} \\
F_{j'k}^{(DPA)} & =   \frac{\hat{\omega}_j'^2}{8\hat{k}_j'^2} \partial_{\rho_j'} \ln \phi_{j'0}\\
f_{j'k}^{(DPA)} &   =(-1)^{j'+1} \left (   \frac{\hat{\omega}_{j'}}{2\hat{k}_{j'}} + \frac{\hat{\omega}_{j'}}{2}  \left| \rho_{j'} \right|\right )
\end{split}
\end{equation}
Note that the DPA has allowed us to obtain explicit expressions for these functions at all distances $z$.

We now consider the asymptotic solutions \eqref{DPABdGFirstOrderSolution} to be the physical DPA modes that we wish to study further. In particular, we consider the DPA ripplons to be perturbations based on these modes, which are converted to the functions $u_{jk}$ and $v_{jk}$ through \eqref{generalformDS}, and then substituted in \eqref{eq:deltaphiuv}. 

\subsection{Dispersion relation for DPA ripplons in the long wavelength limit}
Our goal is to derive a dispersion relation for ripplons that are based on the modes \eqref{DPABdGFirstOrderSolution}.  Note that the expressions \eqref{DPABdGFirstOrderSolutions} satisfy the DPA BdG equations \eqref{eq:DPABdGESigmaDelta} order by order in the expansion for long wavelengths. However, after expanding and truncating to obtain the final asymptotic forms  
\eqref{DPABdGFirstOrderSolution}, the DPA BdG equations are no longer satisfied for arbitrary frequencies and wave numbers. It is this property of  \eqref{DPABdGFirstOrderSolution} that allows us to obtain a {\em constraint} on the physically acceptable combinations of $\omega$ and $k$ for our model ripplon, as follows. 

We suspect that a dispersion relation $\omega(k)$ might be obtained directly from a judicious combination of DPA BdG equations, integrated over space. Using the first and third equations of \eqref{eq:DPABdGESigmaDelta} we derive the identity, 
\begin{eqnarray}\label{identitydisp}
& \int_{-\infty}^{0}  & dz \left (\Sigma_{1k}(\mathcal{A}_{1k} + \beta^2)  \Sigma_{1k}  + \Sigma_{2k}(\mathcal{A}_{2k} + 2)  \Sigma_{2k} \right )  \nonumber \\
 + &\int_{0}^{\infty} & dz \left (\Sigma_{1k}(\mathcal{A}_{1k} + 2)  \Sigma_{1k} +  \Sigma_{2k}(\mathcal{A}_{2k} + \beta^2)  \Sigma_{2k} \right ) \nonumber  \\  = &\sum_{j=1}^2 &\smallint_{-\infty}^{\infty} dz \; \hat{\omega}_{j} \Sigma_{jk}\Delta_{jk}  
\end{eqnarray}
We note that the right-hand-side of this identity, which carries the dimension of length, equals minus the reduced (i.e., scaled with the bulk pressure) excess energy per unit area associated with the capillary wave perturbation \cite{PitaString},
\begin{eqnarray}
- \frac{\delta \gamma_k}{2P_0} \equiv - \frac{\hbar  \omega}{2P_0} \sum_{j=1}^2 \smallint_{-\infty}^{\infty} dz \; n_{j0}\, (|u_{jk}|^2-|v_{jk}|^2 ) = 
\sum_{j=1}^2 \smallint_{-\infty}^{\infty} dz \; \hat{\omega}_{j} \Sigma_{jk}\Delta_{jk}  
\end{eqnarray}

\begin{figure}
\centering
\includegraphics[width=0.88\textwidth]{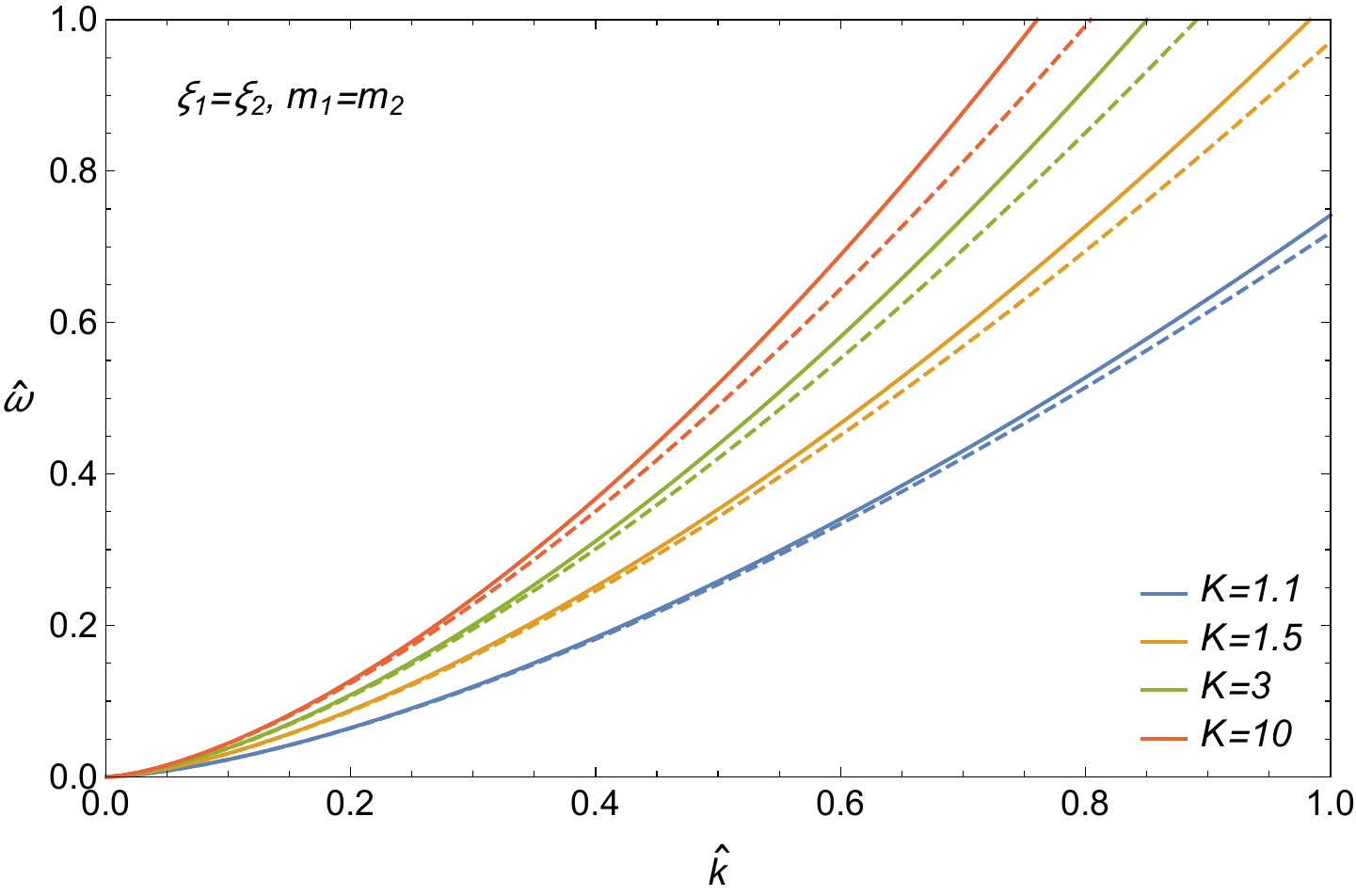}
\caption{Dispersion relation in scaled variables for a symmetric BEC mixture. The asymptotic behavior for long wavelengths is shown (dashed curve) as well as the behavior including a finite-wavelength correction (solid curve). The reduced frequency $\hat \omega$ is determined making use of dispersion relation \eqref{dispersionscaledFullExpanded}. In general the reduced frequency increases when the segregation strength $K$ increases, since then the interfacial tension increases. The upper curves correspond to strong segregation and the lower ones to weak segregation. It is conspicuous that the relative importance of the finite-wavelength correction diminishes as the weak segregation limit ($K\rightarrow 1$) is approached.}
\label{fig:5}
\end{figure}

\begin{figure}
\centering
\includegraphics[width=0.88\textwidth]{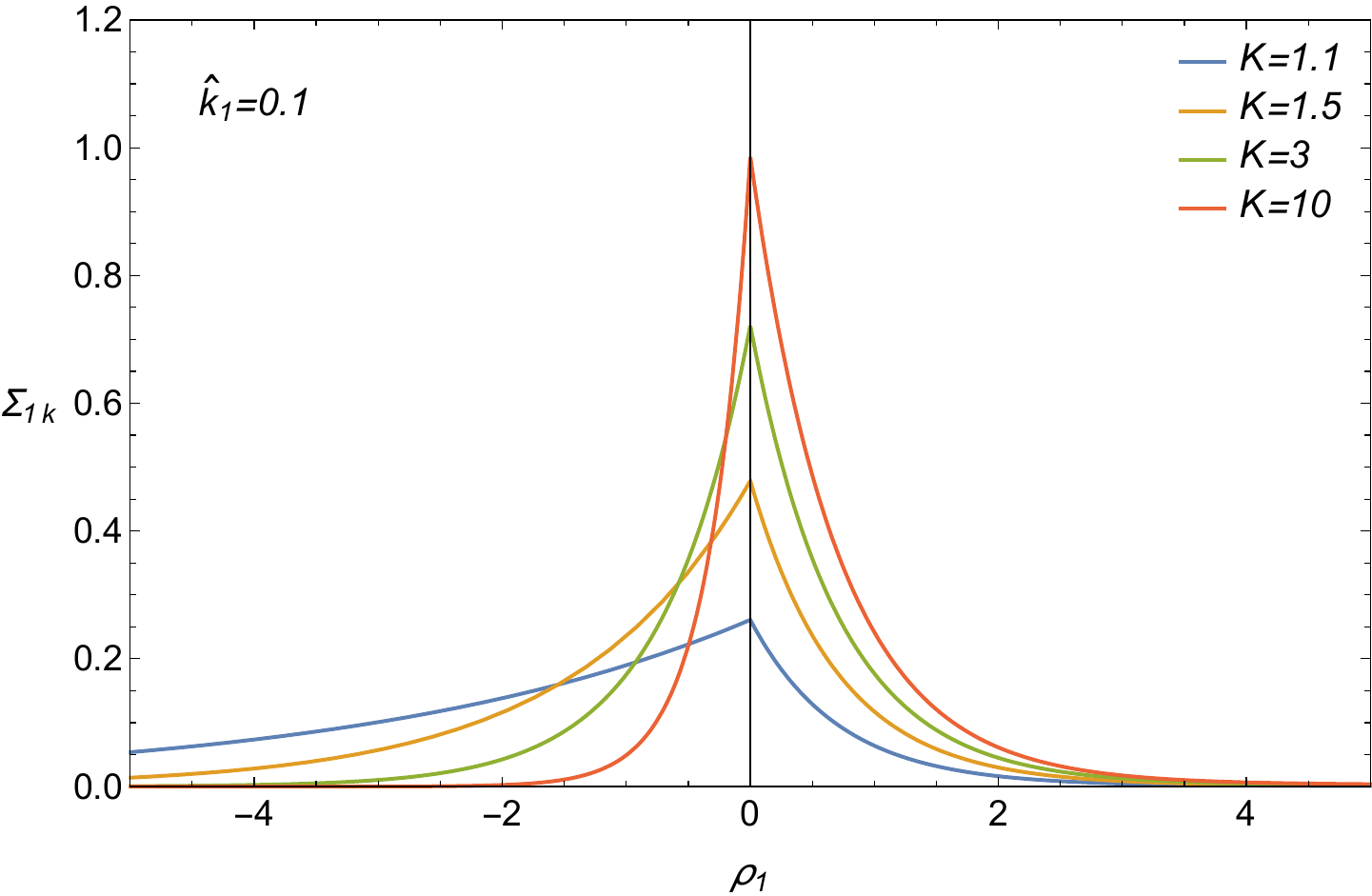}
\caption{Spatial dependence of the analytically calculated $\Sigma_{1k}$-mode, according to  \eqref{DPABdGFirstOrderSolution}, and using the approximation given in the second line for $\Sigma_{jk}$, for various condensate segregation strengths. A symmetric BEC mixture is assumed, so   $\xi_1=\xi_2\equiv \xi$ and consequently $\Sigma_{2k}(\rho)  = - \Sigma_{1k}(-\rho)$, with $ \rho_1 = \rho_2 \equiv \rho$. The reduced wave number is fixed to $\hat{k}_1 = 0.1$ and the reduced frequency $\hat\omega_1 $ is determined using dispersion relation \eqref{dispersionscaledFullExpanded}. The curve with the lowest peak corresponds to reduced interaction strength $K=1.1$ (weak segregation), the two intermediate curves to $K=1.5$ and $K=3$ and the curve with the highest peak to $K=10$ (strong segregation). Note that $\Sigma_{1k}$ is continuous, but its first derivative is discontinuous, at the interface, which is located at $\rho_1=0$. Also note that for $K=3$ the curve is symmetric about the interface. This is a consequence of the symmetry of the analytic forms for $\Sigma$ in \eqref{DPABdGFirstOrderSolution} for $\beta = \sqrt{2}$.}
\label{fig:6}
\end{figure}

\begin{figure}
\centering
\includegraphics[width=0.88\textwidth]{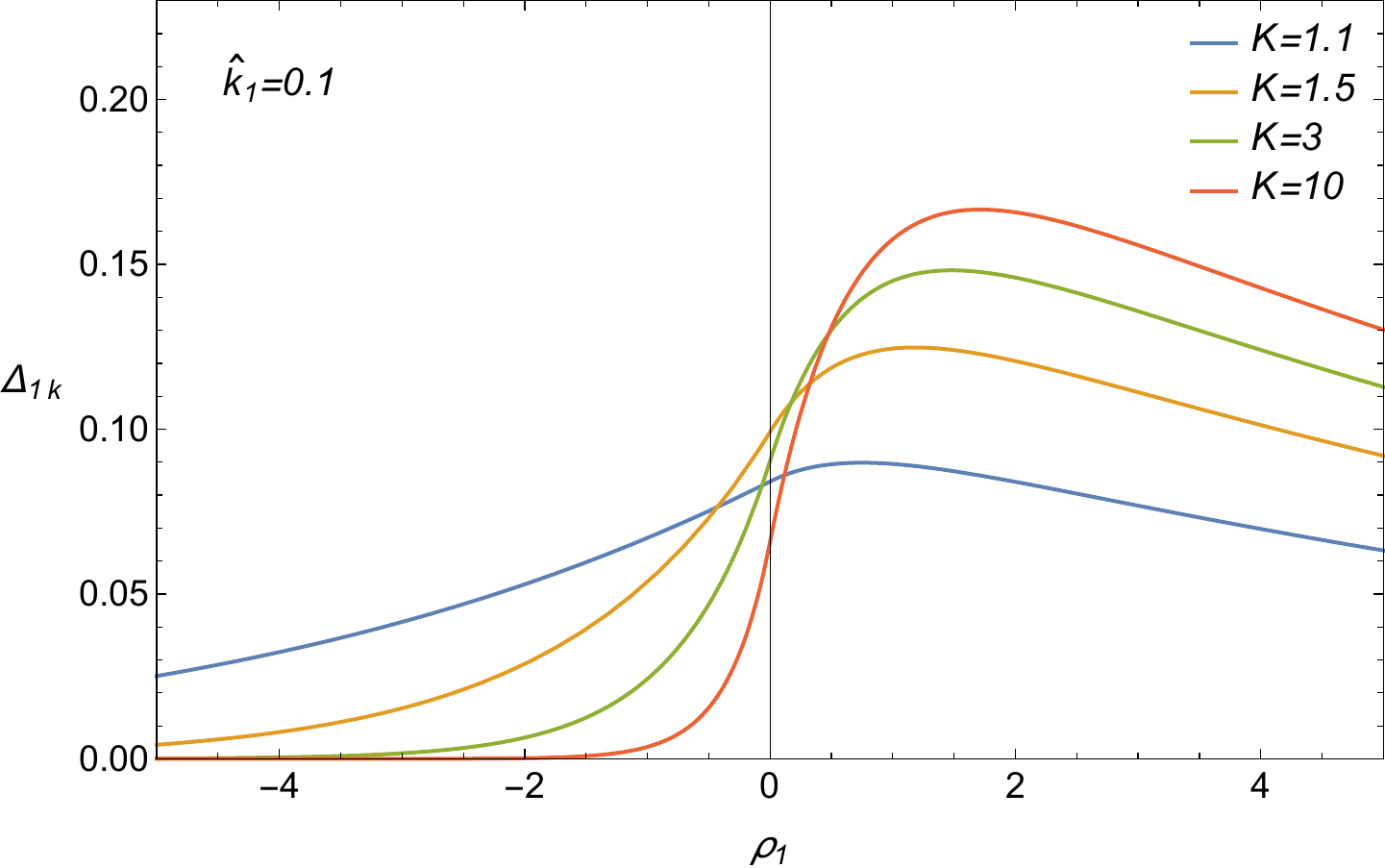}
\caption{Spatial dependence of the analytically calculated $\Delta_{1k}$-mode, according to \eqref{DPABdGFirstOrderSolution}, and using the approximation given in the second line for $\Delta_{j'k}$, for various condensate segregation strengths. A symmetric BEC mixture is assumed, so   $\xi_1=\xi_2\equiv \xi$ and consequently $\Delta_{2k}(\rho) = - \Delta_{1k}(-\rho)$, with $ \rho_1 = \rho_2 \equiv \rho$. The reduced wave number is fixed to $\hat{k}_1 = 0.1$ and the reduced frequency $\hat\omega_1 $ is determined using dispersion relation \eqref{dispersionscaledFullExpanded}. The curve with the mildest slope at the interface corresponds to reduced interaction strength $K=1.1$ (weak segregation), the two intermediate curves to $K=1.5$ and $K=3$ and the curve with the steepest slope at the interface to $K=10$ (strong segregation). Note that $\Delta_{1k}$ and its first derivative are continuous at the interface.}
\label{fig:7}
\end{figure}
 
\begin{figure}
\centering
\includegraphics[width=0.88\textwidth]{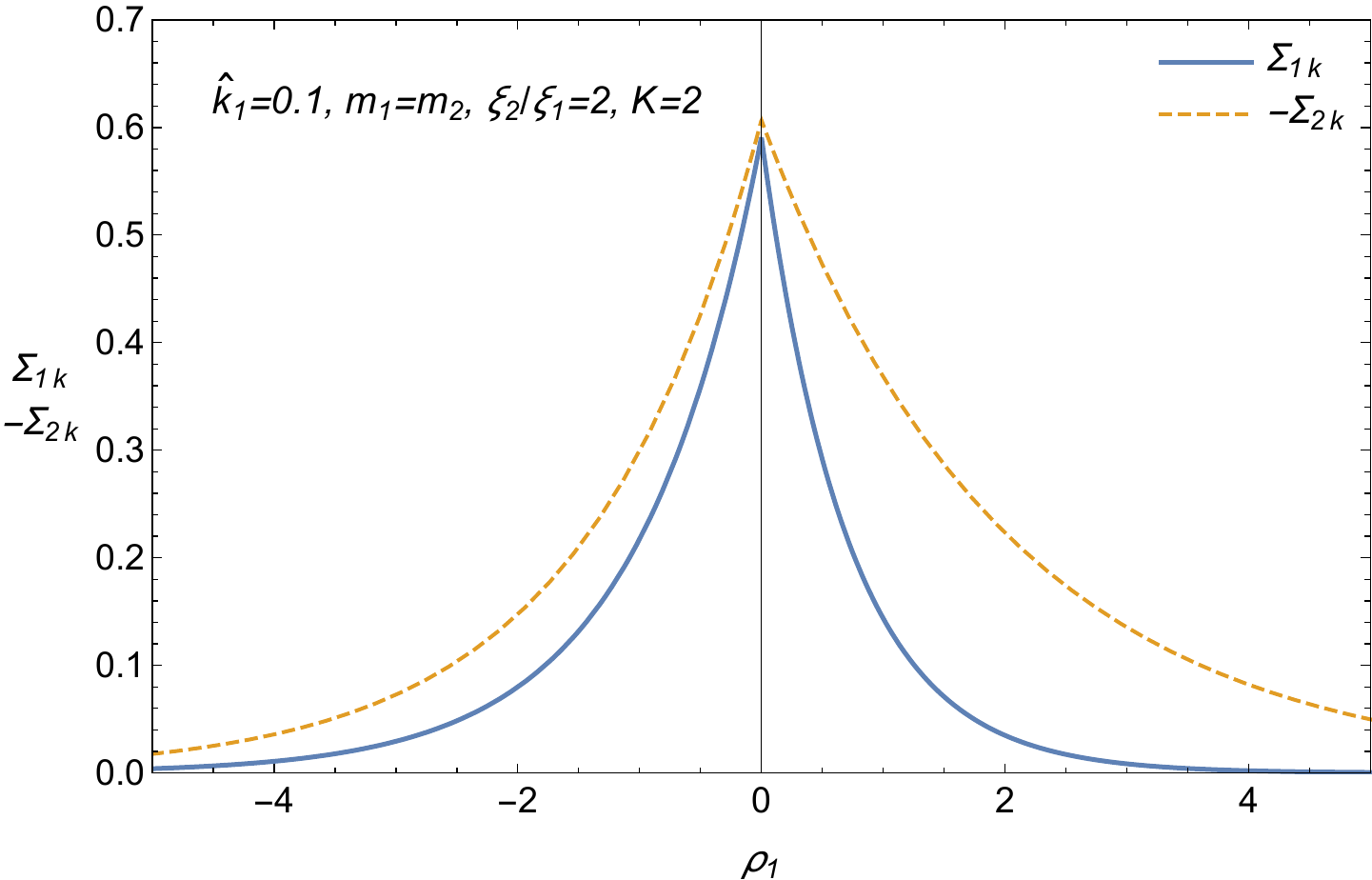}
\caption{Spatial dependence of the analytically calculated $\Sigma_{1k}$ and $-\Sigma_{2k}$, again according to \eqref{DPABdGFirstOrderSolution}, for $K=2$. An asymmetric BEC mixture is assumed, with  $\xi_2=2\xi_1$. The  wave number is fixed through $\hat{k}_1 = 0.1$ and the reduced frequencies are determined using dispersion relation \eqref{dispersionWithCorrectionHandy} and the identity $\hat \omega_2/\hat \omega_1 = (\xi_2/\xi_1)^2$.}
\label{fig:8}
\end{figure}

\begin{figure}
\centering
\includegraphics[width=0.88\textwidth]{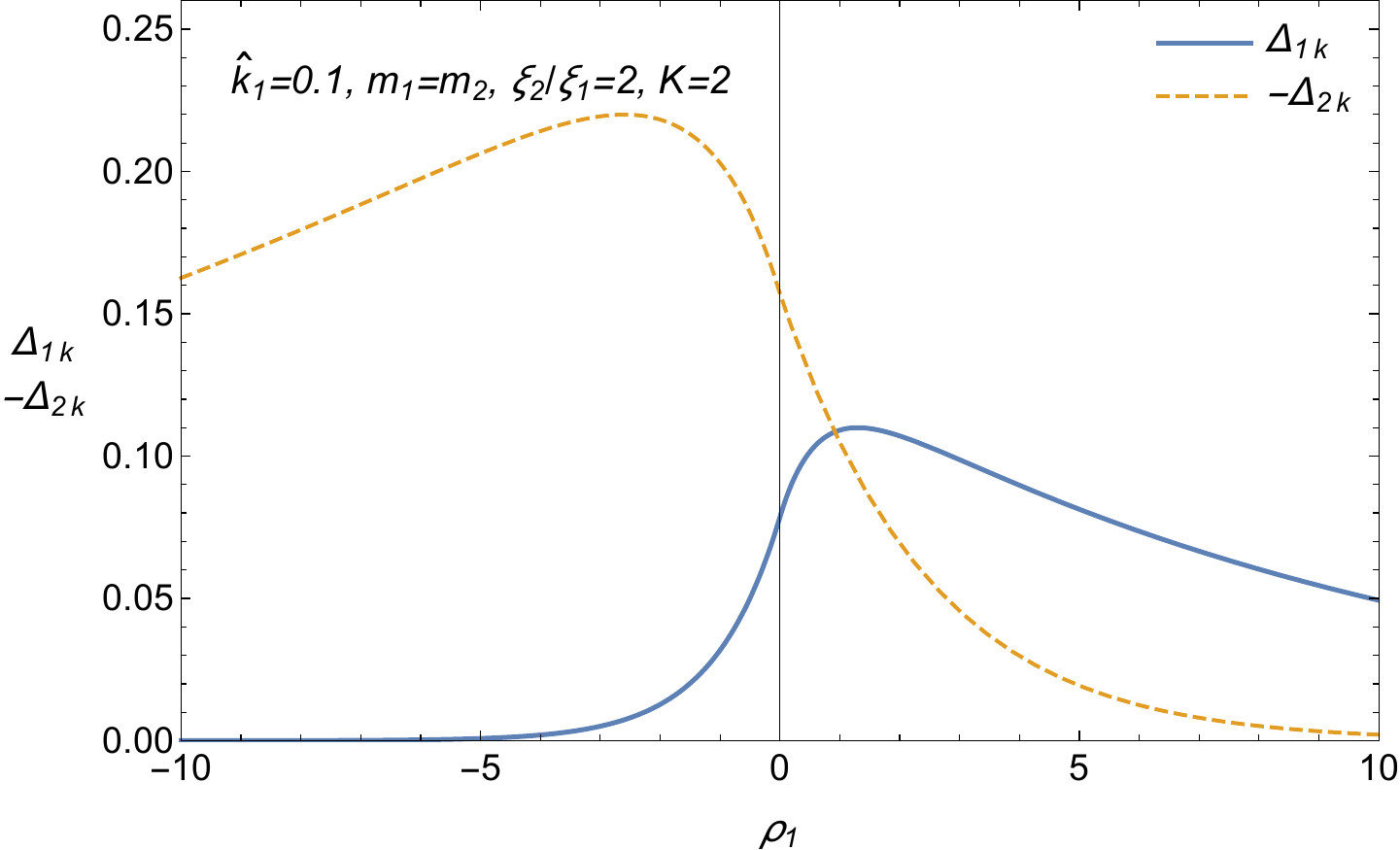}
\caption{Spatial dependence of the analytically calculated $\Delta_{1k}$ and $-\Delta_{2k}$, according to \eqref{DPABdGFirstOrderSolution}, for $K=2$. An asymmetric BEC mixture is assumed, with  $\xi_2=2\xi_1$. The  wave number is fixed through $\hat{k}_1 = 0.1$ and the reduced frequencies are determined using dispersion relation \eqref{dispersionWithCorrectionHandy} and the identity $\hat \omega_2/\hat \omega_1 = (\xi_2/\xi_1)^2$. }
\label{fig:9}
\end{figure}

Now, for the family of modes \eqref{DPABdGFirstOrderSolution} that define the DPA ripplons that we are interested in, the identity \eqref{identitydisp} is not satisfied for arbitrary $\omega$ and $k$, but only for those pairs that satisfy a constraint. Indeed, if we use the asymptotic solutions \eqref{DPABdGFirstOrderSolution},
the identity \eqref{identitydisp} readily leads to the following relationship between the frequencies and the wave numbers, 
\begin{eqnarray}\label{dispersionscaledFull}
\frac{\hat{\omega}_{1}^2 }{\xi_1^2} \;J(\beta) \left (1+  \frac{\hat{\omega}_1^2}{8\hat{k}_1^2} \right ) + \frac{ \hat{\omega}_{2}^2 }{\xi_2^2}\; J(\beta)\left (1+  \frac{\hat{\omega}_2^2}{8\hat{k}_2^2} \right )\nonumber \\ \newline   =  \frac{\hat{k}_1^3}{\xi_1^2} \left (1+ \frac{\hat{\omega}_1^2}{8\hat{k}_1^2} \right )^2+ \frac{\hat{k}_2^3}{\xi_2^2} \left (1+ \frac{\hat{\omega}_2^2}{8\hat{k}_2^2} \right )^2, 
\end{eqnarray}
which reduces, to leading order for long wavelengths, to
\begin{equation}\label{dispersionscaled}
\frac{\hat{\omega}_{1}^2 }{\xi_1^2} +\frac{ \hat{\omega}_{2}^2 }{\xi_2^2} \sim J(\beta)^{-1} \left (\frac{\hat{k}_1^3}{\xi_1^2} + \frac{\hat{k}_2^3}{\xi_2^2} \right ), 
\end{equation}
or, after unscaling the variables, and using \eqref{Jdefactual},
\begin{equation}\label{dispersion}
\omega^2 \sim  \frac{2\sqrt{2}\, \beta }{ \sqrt{2}\,+\beta}  \;\frac{P_0 \,(  \xi_1 +\xi_2)}{m_1n_{10} + m_2n_{20}}  \, k^3
\end{equation}
This is a physically well behaved dispersion relation. Firstly, in the context of BEC mixtures and also far beyond, it is well established that the frequency of capillary waves possesses a leading $k^{3/2}$ dependence on the wave number.  Secondly, specifically for BEC mixtures, the ripplon dispersion takes the form
\begin{equation}
\omega \sim  \sqrt{\frac{\gamma_{12}}{m_1n_{10} + m_2 n_{20}}} \;k^{3/2},
\end{equation}
in the long wavelength limit \cite{c9}. Note that this is formally identical to the dispersion relation for surface waves in classical fluids, known since long \cite{Lamb}.
Here, $m_1 n_{10} + m_2 n_{20}$ is the sum of the condensate mass densities in bulk and $\gamma_{12}$ is the interfacial tension, which takes the following simple analytic form within the static DPA model studied previously \cite{c5},
\begin{equation}\label{eq:gamma12DPA}
\gamma_{12}^{(\rm{DPA})} =  \frac{2 \sqrt{2}\,\beta}{\sqrt{2}+\beta} P_0 (\xi_1 + \xi_2) = \hat\gamma_{12}^{(\rm{DPA})}P_0 (\xi_1 + \xi_2) 
\end{equation}
The interfacial tension implied by \eqref{dispersion} is identical to this expression. We conclude that the dynamic DPA is fully consistent with the static DPA.

A higher-order correction to the dispersion relation can be derived from \eqref{dispersionscaledFull}. The correction in $\omega^2$ is of order $k^4$. For the symmetric case ($\omega_1 = \omega_2 = \omega $ and $\xi_1 = \xi_2 = \xi$) the expression \eqref{dispersionscaledFull} reduces to  
\begin{eqnarray}\label{dispersionscaledFullExpanded}
\hat{\omega}^2   =   \frac{2 \sqrt{2}\,\beta}{\sqrt{2}+\beta}\; \hat{k}^3 \left (1+\frac{ \beta \hat{k}}{2\sqrt{2}\,(\sqrt{2}+\beta)}   + {\cal O}(\hat k^2) \right ),
\end{eqnarray}
with $\hat k = k \xi$.  From this expression the character of the first correction to the long wavelength behavior can be clearly seen. Note that the correction vanishes in the weak segregation limit, relative to the leading term, since it contains an extra factor $\beta = \sqrt{K-1}$. This is a remarkable prediction, which may well survive beyond the DPA, because the DPA is a fairly good approximation for weak segregation.  The dispersion relation including the finite-wavelength correction is shown in Fig.5 for various segregation strengths.

It is straightforward to derive the correction  for the general (asymmetric) case. The result, after unscaling the variables,  is 
\begin{equation}\label{dispersionWithCorrection}
\omega^2 =    \frac{ P_0 (\xi_1 +\xi_2)\,k^3}{  J(\beta)(m_1n_{10} + m_2 n_{20} )}   \left ( 1 +  \beta  k  \,  \frac{  (m_1^2n_{10}^2 - m_2^2n_{20}^2)(\xi_1 -\xi_2) + 2m_1m_2n_{10}n_{20} (\xi_1+ \xi_2) }{ 2\sqrt{2}\,(\sqrt{2}+\beta) (m_1n_{10} + m_2 n_{20})^2 }   + {\cal O}( k^2) \right )
\end{equation}
In scaled variables this can be written in the following handy form,
\begin{equation}\label{dispersionWithCorrectionHandy}
\hat \omega_1^2 =    \frac{ 1 + s}{1+s^2}\,\frac{\hat k_1^3}{J(\beta)}\left ( 1 + \frac{\beta \, \hat k_1}{2\sqrt{2}\,(\sqrt{2}+\beta)}
\left ( 1 + s - 2 \frac{s+s^4}{(1+s^2)^2}\right ) \right ), 
\end{equation}
with $s \equiv \xi_2/\xi_1$.

In the Figures 6-9 we show examples of $\Sigma_{1k}$-modes and $\Delta_{1k}$-modes that are analytically calculated using the asymptotic expressions \eqref{DPABdGFirstOrderSolution} valid for long wavelengths. The dispersion relation \eqref{dispersionWithCorrectionHandy} has been used to determine the frequencies for given wave numbers.

\subsection{Interface deformation, ripplon amplitude enhancement and segregation modulation}   
We are now in a position to present the explicit analytic form of the (moduli of the) condensate order parameters undergoing the capillary wave perturbation of the interface. Recalling \eqref{eq:perturbedphi1d}, we have obtained
\begin{equation}\label{eq:perturbedphi1drecall}
|\phi_j (\rho_j, \chi_j, \tau_j) | = \phi_{j0} (\rho_j) - \varrho_{j0}^{(R)} \partial_{\rho_j} \phi_{j0} (\rho_j)   \sin(kx-\omega t)- \phi_{j0} (\rho_j) \varrho_{j0}^{(R)}\frac{\hat{\omega}_j^2}{8\hat{k}_j^2} \partial_{\rho_j} \ln \phi_{j0} (\rho_j)\sin(kx-\omega t),
\end{equation}
to first order in the amplitude $\varrho_{j0}^{(R)}$ and for long wavelengths. The second term in the right-hand-side is the rigid shift contribution and the last term is the effect of the function $F$. We now regroup terms to obtain
\begin{eqnarray}\label{eq:perturbedphi1drecallsim}
|\phi_j (\rho_j, \chi_j, \tau_j) |&= &\phi_{j0} (\rho_j) - \varrho_{j0}^{(R)} \partial_{\rho_j} \phi_{j0} (\rho_j)(1   +  \frac{\hat{\omega}_j^2}{8\hat{k}_j^2}  ) \sin(kx-\omega t) \nonumber \\ 
&\approx &\phi_{j0} \left(\rho_j- \varrho_{j0}^{(R)} (1   + \frac{\hat{\omega}_j^2}{8\hat{k}_j^2}  ) \sin(kx-\omega t)\right )
\end{eqnarray}
The last of these forms reveals that a deformation of the interface occurs, caused by two condensate-specific shifts that are independent of $z$ and therefore still rigid.  The two shifts are in phase with each other, but may have different amplitudes (except for a symmetric mixture in which case the shift amplitudes are equal). Therefore, the leading-order interface deformation is, {\em i)} a simple {\em enhanced rigid shift} for symmetric mixtures, and {\em ii)} an enhanced rigid shift  combined with a {\em modulation} of the mutual penetration of the condensates. In other words, in the generic case the bare capillary wave is enhanced and decorated with a modulation of the segregation, apparent as a modulation of the overlap of the order parameters at the interface. This overlap is in principle measurable because it corresponds to the first derivative, with respect to the reduced interaction $K$, of the grand potential \cite{BertJoseph}. 

We proceed to calculate the interface displacement $\zeta(x,t)$ using the foregoing result and the intersection criterion \eqref{eq:intersection}, again to first order in $\varrho_{j0}^{(R)}$. We obtain
\begin{equation}\label{IDFprelim}
\zeta(x,t) = \zeta_{0}^{(R)} 
\left ( 1+ \frac{1}{16}\left ( \frac{\hat{\omega}_1^2}{\hat{k}_1^2} + \frac{\hat{\omega}_2^2}{\hat{k}_2^2}
+ (\frac{\hat{\omega}_1^2}{\hat{k}_1^2} - \frac{\hat{\omega}_2^2}{\hat{k}_2^2})
\frac{
\xi_2\partial_{\rho_1} \phi_{10}(0) +\xi_1\partial_{\rho_2} \phi_{20}(0)}
{\xi_2\partial_{\rho_1} \phi_{10}(0) -\xi_1\partial_{\rho_2} \phi_{20}(0)} \right ) \sin(kx-\omega t) \right )
\end{equation}
This expression highlights the symmetric part, corresponding to the average enhanced rigid shift, and the antisymmetric part, associated with the difference of the enhanced rigid shifts of the individual condensates. Regrouping the contributions and reading off the derivatives at $z=0$ from \eqref{eq:stationarysolutionsDPAGPE} leads to the compact result,
\begin{equation}\label{IDF}
\zeta(x,t) = \zeta_{0}^{(R)} \left ( 1+ \frac{\xi_2c_1^{-2}+\xi_1c_2^{-2}}{\xi_1+\xi_2}\frac{\omega^2}{4k^2}\right ) \sin(kx-\omega t),
\end{equation}
where we have used $\hat{\omega}_j/\hat{k}_j = \sqrt{2}\,\omega/(c_j\,k)$, with $c_j$ the sound velocity of condensate $j$ in bulk,
\begin{equation}\label{sound}
c_j\equiv \sqrt{g_{jj}n_{j0}/m_j} = \hbar/(\sqrt{2}\,m_j \xi_j)
\end{equation}
In the symmetric case ($m_1=m_2$ and $\xi_1=\xi_2$) there is only one sound velocity, $c$, and the interface displacement simplifies to
\begin{equation}\label{IDFsym}
\zeta(x,t) = \zeta_{0}^{(R)} \left ( 1+ \frac{\omega^2}{4c^2k^2}\right ) \sin(kx-\omega t),
\end{equation}

\begin{figure}
\centering
\includegraphics[width=0.88\textwidth]{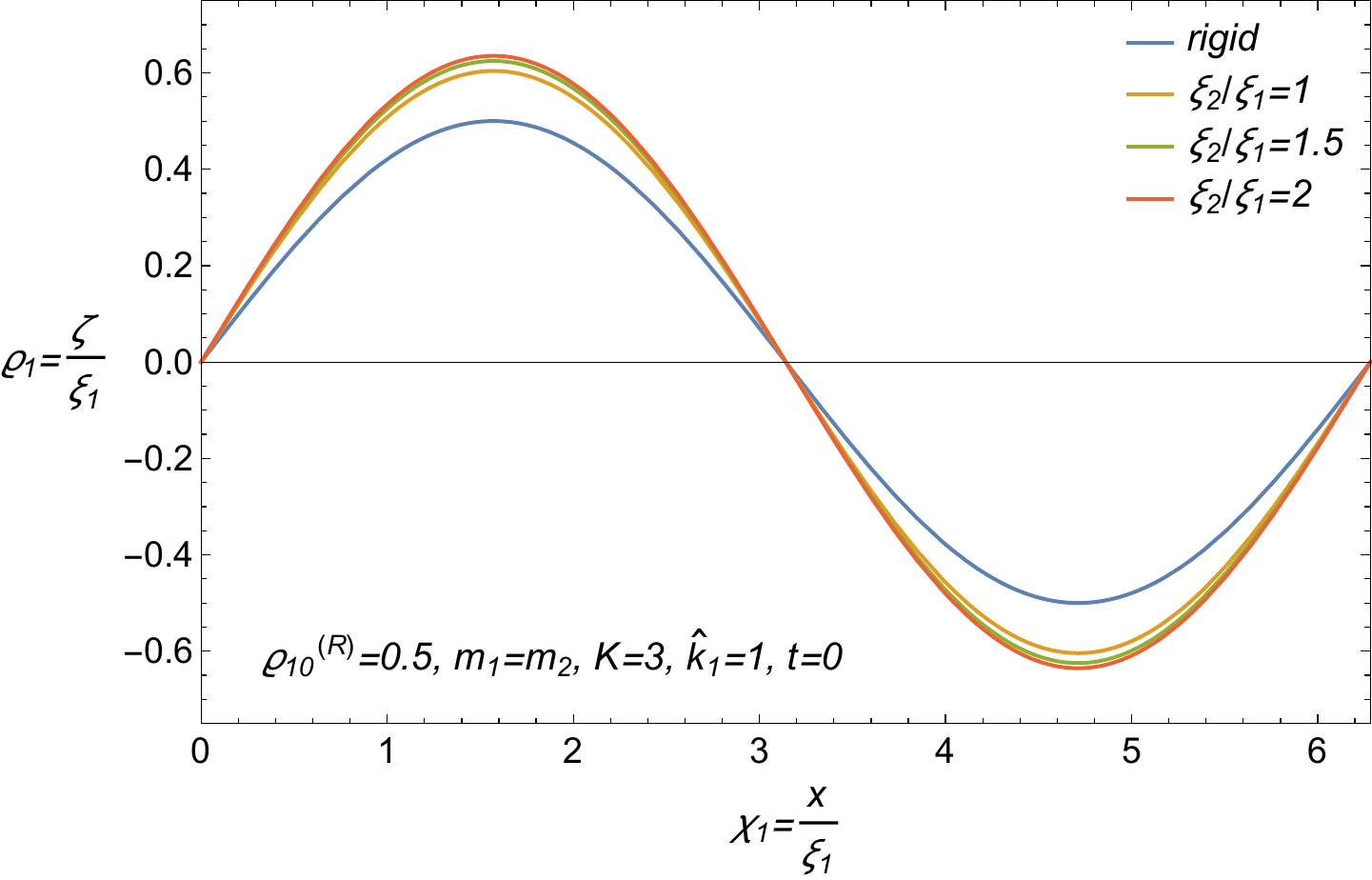}
\caption{Dimensionless interface displacement versus reduced parallel distance along the interface, calculated according to \eqref{IDF}. The amplitude enhancement relative to the bare (i.e., $F_j=0$; ``rigid", innermost curve) rigid-shift displacement depends mainly on the ratio of phase velocity and sound velocity and increases weakly with increasing healing length asymmetry.  }
\label{fig:10}
\end{figure}

\begin{figure}
\centering
\includegraphics[width=0.88\textwidth]{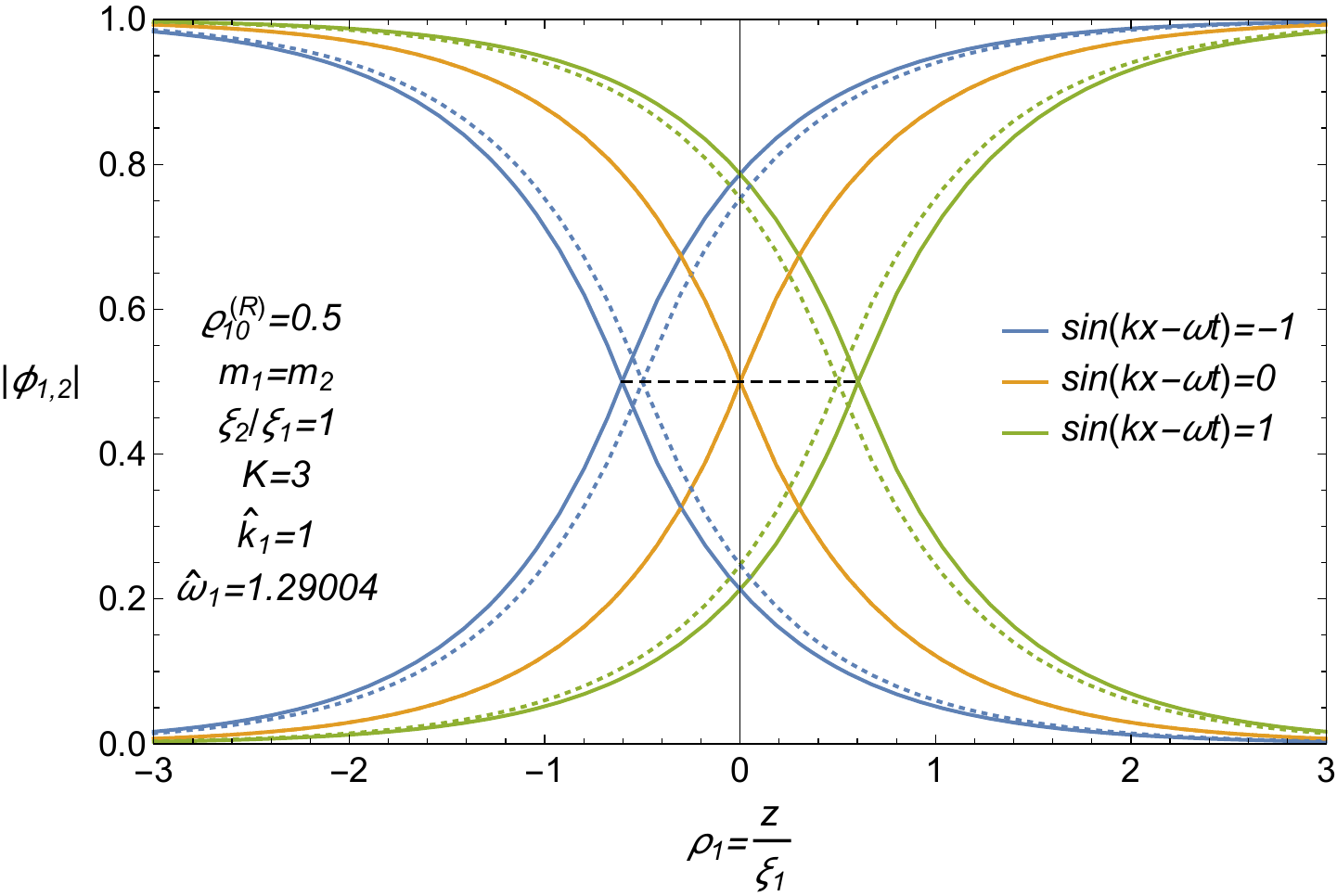}
\caption{Temporal snapshots of the interface profiles illustrating the passing of the capillary wave: symmetric mixture. The actual deformed interface is shown (enhanced rigid shift; solid lines) together with the undeformed profiles (bare rigid shift; dashed lines). Since the mixture is symmetric only the amplitude enhancement described by \eqref{IDF} can be seen. There is no midpoint density modulation. The horizontal dashed segment measures the amplitude enhancement, which obeys  \eqref{IDFsym}.}
\label{fig:11}
\end{figure}

\begin{figure}
\centering
\includegraphics[width=0.88\textwidth]{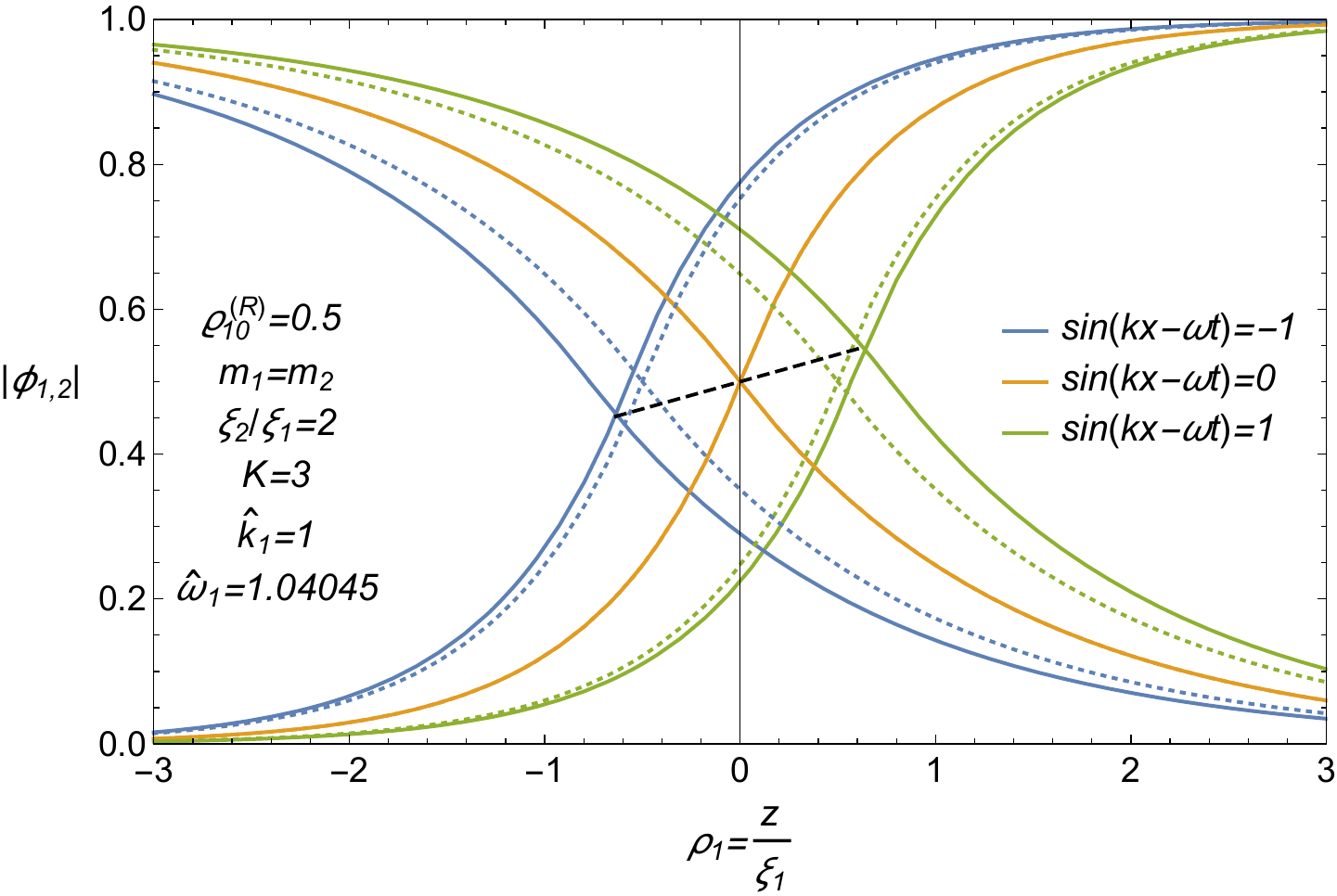}
\caption{Temporal snapshots of the interface profiles illustrating the passing of the capillary wave: (generic) asymmetric mixture. The actual deformed interface is shown (enhanced rigid shift and segregation modulation; solid lines) together with the undeformed profiles (bare rigid shift; dashed lines). The mixture is chosen asymmetric so that not only the amplitude enhancement described by \eqref{IDF} can be seen, but also the midpoint density modulation \eqref{midpointModul}. The inclined dashed segment quantifies the deformation: its horizontal projection measures the enhanced amplitude, which obeys  \eqref{IDF}, while its vertical projection measures the midpoint density range, which obeys \eqref{midpointModul}, to first order in $\zeta_{0}^{(R)}$.}
\label{fig:12}
\end{figure}

We already mentioned that, for generic asymmetric mixtures, the interface deformation comprises a modulation of the interface midpoint density. This phenomenon implies a periodic variation of the extent to which phase segregation takes place. A decrease in the midpoint density is an apparent strengthening of segregation between the condensates, while a midpoint density increase corresponds to more mutual penetration and thus a weakening of segregation. Performing a first-order Taylor expansion of the common midpoint densities \eqref{eq:intersection} about the unperturbed midpoint value $\phi_{j0}(0)$, we find that the segregation modulation is described by the following result, recalling $\varrho_j = \zeta/\xi_j$,
\begin{equation}\label{midpointModul}
|\phi_1(\rho_1=\varrho_1,\chi_1, \tau_1)|  = |\phi_2(\rho_2=\varrho_2,\chi_2, \tau_2)| = \phi_{j0}(0)\left ( 1 +  \frac{\zeta_{0}^{(R)}\beta}{\xi_1+\xi_2} \left  ( \frac{1}{c_2^2}-\frac{1}{c_1^2}  \right )\frac{\omega^2}{4k^2} \sin(kx-\omega t)\right ),
\end{equation}
where $\phi_{j0}(0) =  \sqrt{2}/(\sqrt{2} + \beta)$ is the unperturbed midpoint value. Note that the modulation is antisymmetric for reflection about $z=0$ (exchange of condensates) and vanishes in the symmetric case $c_1=c_2$ of equal sound velocities. It is noteworthy that the modulation vanishes in the weak segregation limit. The predicted  exponent, i.e., the power of $K-1$, takes the value $1/2$, which is the same exponent as for the interfacial tension.  

Figures 11 and 12 illustrate the interfacial order parameter profiles with quarter period time intervals during the passing of the capillary wave. Fig.11 deals with a symmetric case, for which the interface deformation consists merely of an enhancement of the rigid shift (solid lines) as compared with the bare rigid shift associated with $F_j=0$ (dashed lines). Fig.12 exemplifies the generic case $\xi_1 \neq \xi_2$, for which also a modulation of the midpoint density occurs. Finally, Fig.13 illustrates the interfacial density modulation for asymmetric mixtures, which increases with increasing asymmetry.

\begin{figure}
\centering
\includegraphics[width=0.88\textwidth]{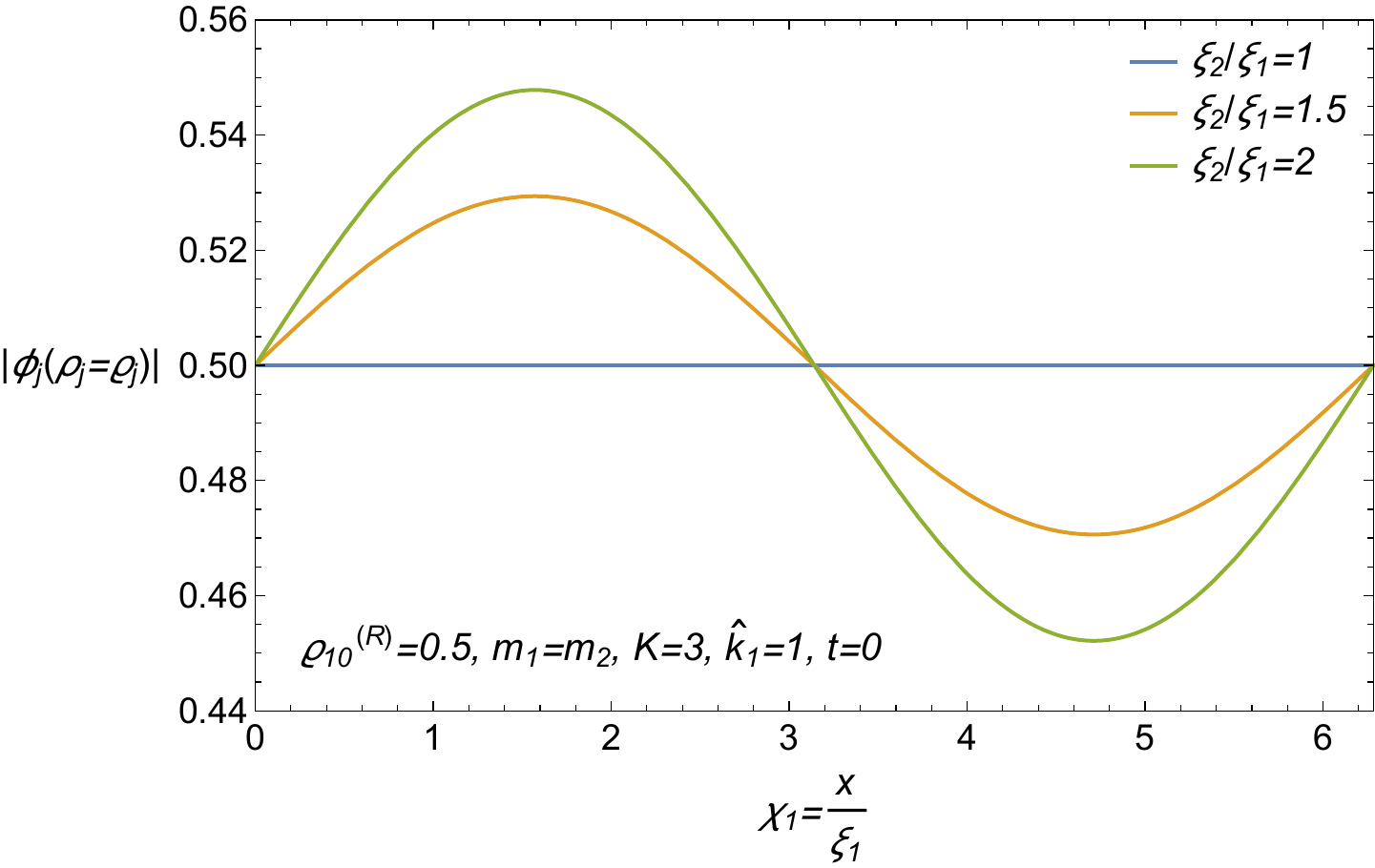}
\caption{The modulation of the condensate  midpoint density for asymmetric BEC mixtures, as described by \eqref{midpointModul}. A single spatial period is shown.}
\label{fig:13}
\end{figure}

For completeness we verify whether the interface displacement velocity is consistent with the phase fluctuation of each component, as expressed by \eqref{vconsistent}. For this it suffices to use \eqref{DPAofFandf} for the functions $f_{jk}$ and just the leading term in \eqref{IDF}. We obtain, on the one hand,
\begin{equation}
v_{jz}=\partial _t \zeta = - \omega \zeta_0^{(R)}\cos(kx-\omega t) + ...,
\end{equation}
and, on the other hand, 
\begin{equation}
v_{jz}= \frac{\hbar}{m_j} (\nabla \theta_j)_z = -\omega\zeta_0^{(R)}\cos(kx-\omega t) + ... = \frac{\hbar}{m_{j'}} (\nabla \theta_{j'})_z =v_{j'z} 
\end{equation}
Note that all these velocities should be equal and independent of the condensate labels $j$ and $j'$. This is indeed the case.  

\section{Conclusions and outlook}
In this paper we have developed a dynamic Double Parabola Approximation, starting from the static DPA model previously studied. To this end, we first derived a DPA to the time-dependent GP equations. From these we derived DPA BdG equations and so defined a model for studying ripplon excitations. We have also shown that these approximations commute: applying a DPA to the original BdG equations derived from the time-dependent GP equations, leads to the same model. Within this model we  established explicit mathematical DPA solutions for ripplons for all distances near to and far from the interface and checked that the formalism allows one to obtain correct zero modes consistent with exact relations in GP theory. The accuracy of the DPA relies on the assumption that both $\sqrt{2}/(\sqrt{2}+\sqrt{K-1})$ and $\sqrt{K-1}/(\sqrt{2}+\sqrt{K-1}) $ be small compared to unity. These requirements are, strictly speaking, mutually incompatible since the first parameter is only small for  strong segregation and the second only for weak segregation. However, the fact that both parameters are bounded from above by unity renders the DPA useful in all segregation regimes. The usefulness of the DPA was already established for calculating static properties analytically \cite{c5} and it emerges here once more in the context of time-dependent properties and phenomena.

Using the DPA solutions for the modes in the long wavelength limit, we derived a family of modes and used that to define a physical DPA ripplon. For this type of perturbation we derived a dispersion relation $\omega (k)$ directly from the DPA BdG equations combined in integral form and used as a constraint on allowed values for frequencies and wave numbers. 
In the long wavelength limit the dispersion relation carries the correct wave number dependence $k^{3/2}$ and satisfies the correct mean-field scaling behavior in the weak segregation limit, reproducing the  exponent $1/4$ describing the dependence of the square root of the interfacial tension on the interaction strength difference $K-1$. The leading term in the dispersion relation features the static interfacial tension derived earlier within the DPA \cite{c5}. Therefore, the dynamic DPA  fully embodies the static DPA.  

The next-to-leading term in the dispersion relation, of order $k^{5/2}$, is also derived analytically for all interaction strengths $K$. This term features its own exponent $3/4$, implying that the correction vanishes faster than the leading term for $K \rightarrow 1$. The correction term has to our knowledge not received much attention in the literature, the only exception we know of being a result derived in the strong segregation limit by Mishonov \cite{Misho}. Comparing our correction factor in $\omega^2$, which is 1+ $\sqrt{K-1}\,k \xi/(2\sqrt{2}\,(\sqrt{2}+\sqrt{K-1})) $  in \eqref{dispersionscaledFullExpanded}, for $K \rightarrow \infty$, with Mishonov's correction factor in $\omega^2$,  which we estimated to be $1+0.557 \times 2 \sqrt{2} \,k \xi$, would indicate that his correction is significantly larger than ours. This discrepancy merits further investigation. The deviation from the $k^{3/2}$ dispersion at finite wavelength is a prediction that can be verified experimentally. State-of-the-art experiments of BEC in (quasi-)uniform optical-box traps \cite{Hadzi} would be the appropriate arena for testing this result.

The dynamic DPA model  further predicts a small structural deformation of the interface due to the passing of the capillary wave, which can be calculated analytically. Its magnitude is of order $\omega^2/(kc)^2$, with $\omega/k$ the phase velocity and $c$ the sound velocity. The deformation consists of an enhancement of the amplitude of the wave for all types of mixtures. For generic asymmetric mixtures consisting of condensates with unequal healing lengths $\xi_1 \neq \xi_2$ an additional modulation is predicted of the common value of the condensate densities at the interface. This oscillation of the midpoint density is reminiscent of what would happen in the presence of a longitudinal breather mode. This brings us to new applications of the dynamic DPA to other excitations, e.g., interface phonons, which are postponed to future work.

\begin{acknowledgments}
N.V.T and T.H.P are supported by the Vietnam National Foundation for Science and Technology Development (NAFOSTED) and J.O.I. and C.-Y.L. by FWO Flanders under Grant Nr.FWO.103.2013.09 within the
framework of the FWO-NAFOSTED cooperation. J.O.I. and C.-Y.L. are furthermore supported by  KU Leuven Grant OT/11/063. N.V.T is also supported by the Ministry of Education and Training of Vietnam (Grant B2016-SP2-04). The authors thank Hans Hooyberghs, Mehran Kardar, Todor Mishonov, Lev Pitaevskii and Bert Van Schaeybroeck for extensive discussions.
\end{acknowledgments}

\appendix

\section{Equations in Gross-Pitaevskii theory}\label{sec:GPtheory}

In our derivations we use the following results from the GP theory. The GP potential is
\begin{equation}\label{eq:potential}
\tilde{\mathcal{V}}_{\mathrm{GP}}(\phi_1,\phi_2)= \sum_{j=1}^{2}  \left[-\lvert\phi_j\rvert^2 + \frac{\lvert\phi_j\rvert^4}{2} \right] + K \lvert\phi_1\rvert^2 \lvert\phi_2\rvert^2.
\end{equation}
The GP equations are
\begin{equation}\label{eq:GPE}
i \partial_{\tau_j} \phi_j  = \left[- \nabla_{\mathbf{s}_j}^2 -1 + \lvert\phi_j\rvert^2 + K \lvert\phi_{j'}\rvert^2 \right] \phi_j   \;, \; j=1,2 \; (j \ne j'),
\end{equation}
The BdG equations are
\begin{equation}\label{eq:BdGEdeltaphi}
i \partial_{\tau_j} \delta \phi_j  = \left[- \nabla_{\mathbf{s}_j}^2 -1 + 2 \phi_{j0}^2 + K \phi_{j'0}^2 \right] \delta\phi_j + \phi_{j0}^2 \delta\phi_j^{\ast} + K \phi_{j0}\phi_{j'0} (\delta\phi_{j'} +\delta\phi_{j'}^{\ast}), \; j=1,2 \; (j \ne j'),
\end{equation}
or, by substituting $\delta \phi_j $ with Bogoliubov form \eqref{eq:deltaphi},
\begin{equation}\label{eq:BdGEuv}
\begin{split}
\left(\mathcal{A}_{jk}  -1+ 2 \mathcal{F}_{jj} + K \mathcal{F}_{j'j'} \right)  u_{jk} +\mathcal{F}_{jj} v_{jk} + K \mathcal{F}_{jj'} (u_{j'k}+v_{j'k}) &= \hat{\omega}_j u_{jk} \\
\left(\mathcal{A}_{jk}  -1+ 2 \mathcal{F}_{jj} + K \mathcal{F}_{j'j'} \right)  v_{jk} + \mathcal{F}_{jj} u_{jk} + K \mathcal{F}_{jj'} (u_{j'k}+v_{j'k}) &= -\hat{\omega}_j v_{jk} 
\end{split}
 \;\; \;{\rm with} \; j=1,2 \; (j \ne j'),
\end{equation}
or, in terms of $\Sigma_{jk} $ and $\Delta_{jk} $.
\begin{equation}\label{eq:BdGESigmaDelta}
\begin{split}
\left(\mathcal{A}_{jk}  -1+ 3 \mathcal{F}_{jj} + K \mathcal{F}_{j'j'} \right)  \Sigma_{jk} + 2 K \mathcal{F}_{jj'} \Sigma_{j'k} &= \hat{\omega}_j \Delta_{jk} \\
\left(\mathcal{A}_{jk}  -1+  \mathcal{F}_{jj} + K \mathcal{F}_{j'j'} \right)  \Delta_{jk}  &= \hat{\omega}_j \Sigma_{jk} 
\end{split}
 \;\; \;{\rm with} \; j=1,2 \; (j \ne j'),
\end{equation}
where $\mathcal{A}_{jk} = \hat{k}_j^2- \partial^2_{\rho_j} $ and $\mathcal{F}_{jj'} = \phi_{j0} \phi_{j'0} $.

The solutions in the limit $\hat k \rightarrow 0$ and $\hat \omega \rightarrow 0$ are called the zero modes. It is easy to obtain the $z$-dependence of these modes. They solve
\begin{equation}\label{eq:BdGESigmaDelta}
\begin{split}
\left(- \partial^2_{\rho_j}  -1+ 3 \mathcal{F}_{jj} + K \mathcal{F}_{j'j'} \right)  \Sigma_{j0} + 2 K \mathcal{F}_{jj'} \Sigma_{j'0} &= 0\\
\left(- \partial^2_{\rho_j}  -1+  \mathcal{F}_{jj} + K \mathcal{F}_{j'j'} \right)  \Delta_{j0}  &= 0
\end{split}
  \;\; \;{\rm with} \;j=1,2 \; (j \ne j').
\end{equation}
The second of these equations is just the static GP equation and is solved by
\begin{equation}\label{exactzeroD}
\Delta_{j0} \propto \phi_{j0}, \;\; j=1,2.
\end{equation}
Inserting this result in the second equation and then taking the $z$-derivative of it, suffices to show that the first equation is solved by
\begin{equation}\label{exactzeroS}
\Sigma_{j0} \propto \partial_{\rho_j} \phi_{j0}, \;\; j=1,2.
\end{equation}


\end{document}